\documentclass[12pt]{article}

\usepackage{scicite}

\usepackage{amsmath}
\usepackage{amssymb}
\usepackage{mathtools}
\usepackage{times}
\usepackage{float}
\usepackage{ulem}
\usepackage{soul}

\usepackage{graphicx}
\usepackage{caption}
\usepackage{hyperref}
\usepackage{color}
\usepackage{bm}

\newenvironment{sciabstract}{%
\begin{quote} \bf}
{\end{quote}}

\title{Sub-nanotesla Sensitivity at the Nanoscale with a Single Spin}

\author
{Zhiyuan Zhao$^{1,2\dag}$, Xiangyu Ye$^{1,2\dag}$, Shaoyi Xu$^{1,2}$, Pei Yu$^{1,2}$, Zhiping Yang$^{1,2}$\\Xi Kong$^{3}$, Ya Wang$^{1,2,4}$, Tianyu Xie$^{1,2\ast}$, Fazhan Shi$^{1,2,4,5\ast}$, Jiangfeng Du$^{1,2,4\ast}$\\
\\
\normalsize{$^{1}$CAS Key Laboratory of Microscale Magnetic Resonance and School of Physical Sciences,}\\
    \normalsize{University of Science and Technology of China, Hefei 230026, China}\\
\normalsize{$^{2}$CAS Center for Excellence in Quantum Information and Quantum Physics,}\\
	\normalsize{University of Science and Technology of China, Hefei 230026, China}\\
\normalsize{$^{3}$National Laboratory of Solid State Microstructures and Department of Physics,}\\
	\normalsize{Nanjing University, Nanjing 210093, China}\\
\normalsize{$^{4}$Hefei National Laboratory,}\\
    \normalsize{University of Science and Technology of China, Hefei 230088, China}\\
\normalsize{$^{5}$School of Biomedical Engineering and Suzhou Institute for Advanced Research,}\\
    \normalsize{University of Science and Technology of China, Suzhou 215123, China}\\
\\
\normalsize{$^\dag$These authors contributed equally to this work.}\\
\normalsize{$^\ast$E-mail: xie1021@ustc.edu.cn}\\
\normalsize{$^\ast$E-mail: fzshi@ustc.edu.cn}\\
\normalsize{$^\ast$E-mail: djf@ustc.edu.cn}\\
}

\date{}

\begin{document}

\captionsetup[figure]{labelfont={bf},labelsep=period,name={Fig.}}
\baselineskip24pt

\maketitle

\pagebreak
\begin{sciabstract}
High-sensitivity detection of microscopic magnetic field is essential in many fields.
Good sensitivity and high spatial resolution are mutually contradictory in measurement, which is quantified by the energy resolution limit (ERL).
Here we report that a sensitivity of {0.5} ${\bf{nT/\sqrt{Hz}}}$ at the nanoscale is achieved experimentally {by using nitrogen-vacancy defects in diamond with depths of tens of nanometers.}
The achieved sensitivity is substantially enhanced by integrating with multiple quantum techniques, including real-time-feedback initialization, dynamical decoupling with shaped pulses, repetitive readout via quantum logic.
Our magnetic sensors will shed new light on searching new physics beyond the standard model, investigating microscopic magnetic phenomena in condensed matters, and detection of life activities at the sub-cellular scale.
\end{sciabstract}

\section*{Introduction}

Along with the advances in science, more investigations are focusing on the microscopic regime, and naturally require the detection of magnetic field at the microscale \cite{Lovchinsky2016Science, Safronova2018RMP, Cheong2020npj, Chibotaru2000Nature, Abobeih2019Nature, Frederick2015Cell, Rong2018PRL, Glover2002RPP, Blatter1994RMP, Wishart1991jmb, Sikivie2021RMP}.
For instance, magnetic sensors have been used to search for new interactions beyond the standard model \cite{Safronova2018RMP, Sikivie2021RMP}.
It is inevitable to build a microscopic sensor with an extremely high sensitivity for exploring interactions in the short-range regime \cite{Rong2018PRL, Vinante2021PRL}.
In condensed matters, observations of some nanoscale magnetic phenomena, like antiferromagnetic domains with extremely weak magnetization \cite{Cheong2020npj} and novel behaviors in mesoscopic superconductors \cite{Chibotaru2000Nature, Blatter1994RMP}, also have similar requirements.
Biological applications at the microscale, especially for biochemical reactions at the sub-cellular level and nuclear magnetic resonance imaging of a single cell \cite{Glover2002RPP}, demand both sub-nanotesla sensitivity and sub-micron spatial resolution.

However, the performances of the magnetometers invented so far are still to be improved for the applications above. As the most sensitive magnetic sensor, superconducting quantum interference devices (SQUIDs) can provide sub-femtotesla sensitivity at  the macroscale \cite{Schmelz2016IEEE}, but working at a smaller scale will dramatically worsen its sensitivity, e.g., tens of nT$\rm/\sqrt{Hz}$ at $\sim$ 100 nanometers \cite{Vasyukov2013NatureNano}.
On the other hand, single nitrogen-vacancy (NV) centers in diamond have an extraordinary performance at the nanoscale but still with a poor sensitivity of tens of nT$\rm/\sqrt{Hz}$ at best \cite{Lovchinsky2016Science}.
Other (more than twenty) kinds of magnetometers have similar behaviors, among which the most prominent are optically pumped magnetometers (OPM) \cite{Kominis2003Nature}, Bose-Einstein condensates (BEC) \cite{Vengalattore2007PRL}, and ensemble NV centers \cite{Wolf2015PRX}.
It seems not realistic to perform a high-spatial-resolution detection without impairing the sensitivity.
As an empirical limit, the energy resolution limit $E_R=\hbar$ (ERL) is put forward to quantitatively portrait the contradiction. Up to now, {more than twenty magnetometer technologies} comply with the ERL \cite{Mitchell2020RMP}.

\begin{figure*}\center
\includegraphics[width=1.0\textwidth]{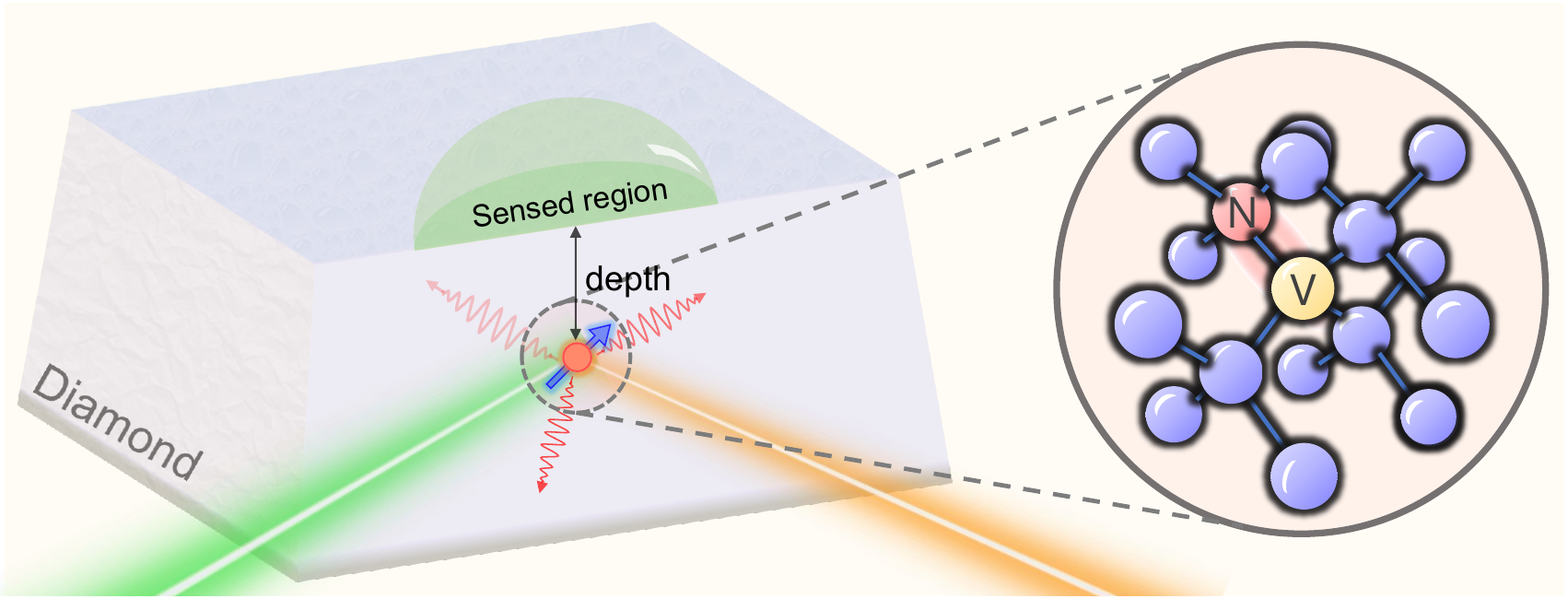}
\caption{\label{setup}\textbf{Experimental system.} Diagram of a single near-surface NV center in diamond.
   The depth of the NV center determines the dimension of the sensed region.
   The 532-nm green laser is used for readout of NV electron spin by collecting fluorescence and mixing charge states, and the 594-nm orange laser is used for readout of charge state in real-time-feedback initialization.
   Inset: the atomic structure of the NV center in diamond lattice.
}
\end{figure*}

{In this work, to overcome the aforementioned problem to some extent, we experimentally achieve the sensitivity of 0.5 $\rm{nT/\sqrt{Hz}}$ for NV centers with depths of more than 30 nm.} The NV center is a point defect that consists of a substitutional nitrogen atom and an adjacent vacancy, as shown in Fig. \ref{setup}, and resides near the diamond surface.
The depth is the minimum distance that a sample to be detected can approach, and thus determines the spatial resolution \cite{Staudacher2013Science, Lovchinsky2016Science}.
In our work, the depth of the NV center is measured by detecting the nuclear magnetic resonance from the proton sample put upon the diamond surface \cite{Pham2016PRB} (see Materials and Methods for more details).
The detected signal mainly originates from the protons at a distance comparable to the NV depth, which justifies taking the depth roughly as the  achieved spatial resolution.

\section*{{Results}}
\subsection*{{Sensitivity optimization}}
By comprehensively considering multiple experimental limitations, including spin decoherence, initialization and readout errors, and duty cycle, the magnetic sensitivity $\eta_B$ for an individual NV center can be accurately formulated as
\begin{equation}
\label{sensitivity}
{\eta_B=\frac{1}{\gamma_e\sqrt{T}}\cdot \frac{1}{C_T F_r F_i}\cdot\sqrt{1+\frac{T_{ir}}{T}}}
\end{equation}
{where $\gamma_e$ is the gyromagnetic ratio of the electron spin, $T$ is the time for phase accumulation, and $C_T$ is the remained spin coherence at $T$.}
The initialization and the readout are both imperfect with $F_i$ and $F_r$ denoting their respective fidelities, and occupy considerable time $T_{ir}$ leading to a reduced duty cycle. 
By controlling the temperature stability to sub-millikelvin levels {(see Fig. S11)}, the magnetic field is sufficiently stable so that no calibration of the magnetic field is required during the experiment, and therefore the experimental duty cycle does not need to take into account any other time such as magnetic field calibration.
For the near-surface NV center that is exposed to the noise of the diamond surface, the coherence time is shortened dramatically down to tens of microseconds \cite{Degen2012PRB}, compared with several milliseconds for NV centers deep inside a bulk diamond \cite{Herbschleb2019NC}.
For the initialization of the NV center by 532-nm laser illumination, there exist two charge states, of which the useful NV$^-$ state occupies $\sim$70\% and the left is the useless NV$^0$ state \cite{Xie2021SciAd}.
The percentage of the NV$^-$ state would be further reduced with shallower NV centers \cite{Bluvstein2019PRL}.  
On the optical readout of the NV spin state, the readout fidelity $F_r$ is rather low, $\sim$3\% for typical fluorescence collection efficiencies\cite{Lovchinsky2016Science}. Substituting the values above into the Eq. \ref{sensitivity} gives the sensitivity roughly 0.1 \textmu T$\rm/\sqrt{Hz}$ for typical 10-nm-deep NV centers {(if the coherence time is chosen to be 50 \textmu$\rm s$, $C_T=1/e$)}.

Improving the NV spin coherence is of paramount importance. Apart from directly enhancing the sensitivity, long coherence times also enable the use of some time-consuming quantum techniques without dramatically decreasing duty cycles. In order to attenuate the adverse effect from the surface noise, we choose to use NV centers with depths of 10-100 nm combined with the technique of dynamical decoupling \cite{Du2009Nature}. Using an XY16-512 sequence, the coherence time is extended from 146 $\pm$ 5 \textmu s (Hahn echo) to 2.0 $\pm$ 0.2 ms for the NV center with a depth of 31.7 $\pm$ 1.1 nm. Besides, the magnetic noise from $^{13}$C nuclear spins in diamond lattice is removed by $^{12}$C isotope purification.

With millisecond-scale spin coherence, two quantum techniques, real-time feedback for NV negative state preparation \cite{Xie2021SciAd} and repetitive readout via quantum logic \cite{Lovchinsky2016Science}, can be integrated into the measurement sequence to improve initialization and readout fidelities.
Through hundreds of real-time feedback loops (the first block of Fig. \ref{sequence}a), the NV$^-$ state can be picked out with a 99\% success rate \cite{Xie2021SciAd,charge_2020}, but too many loops will degrade the sensitivity. 
After the sensitivity is optimized, the initialization fidelity of NV$^-$ state is better than 92\% (see Materials and Methods).
High-fidelity readout is realized by transferring the NV spin state to the nuclear spin of the adjacent nitrogen atom through a SWAP gate\cite{Liang2009Science,Neumann2010Science}, then followed by thousands of readouts of the nuclear spin (the last two blocks of Fig. \ref{sequence}a).
The nondestructive nature of the nuclear spin after every readout is ensured by a high magnetic field (7662 G in our setup). 
{The number of readout cycles is also optimized}, and executing 2500 cycles (1.44 ms) gives a fidelity of roughly 84\%.

\subsection*{{Magnetic field sensing and sensitivity}}
\begin{figure*}\center
\includegraphics[width=1.0\textwidth]{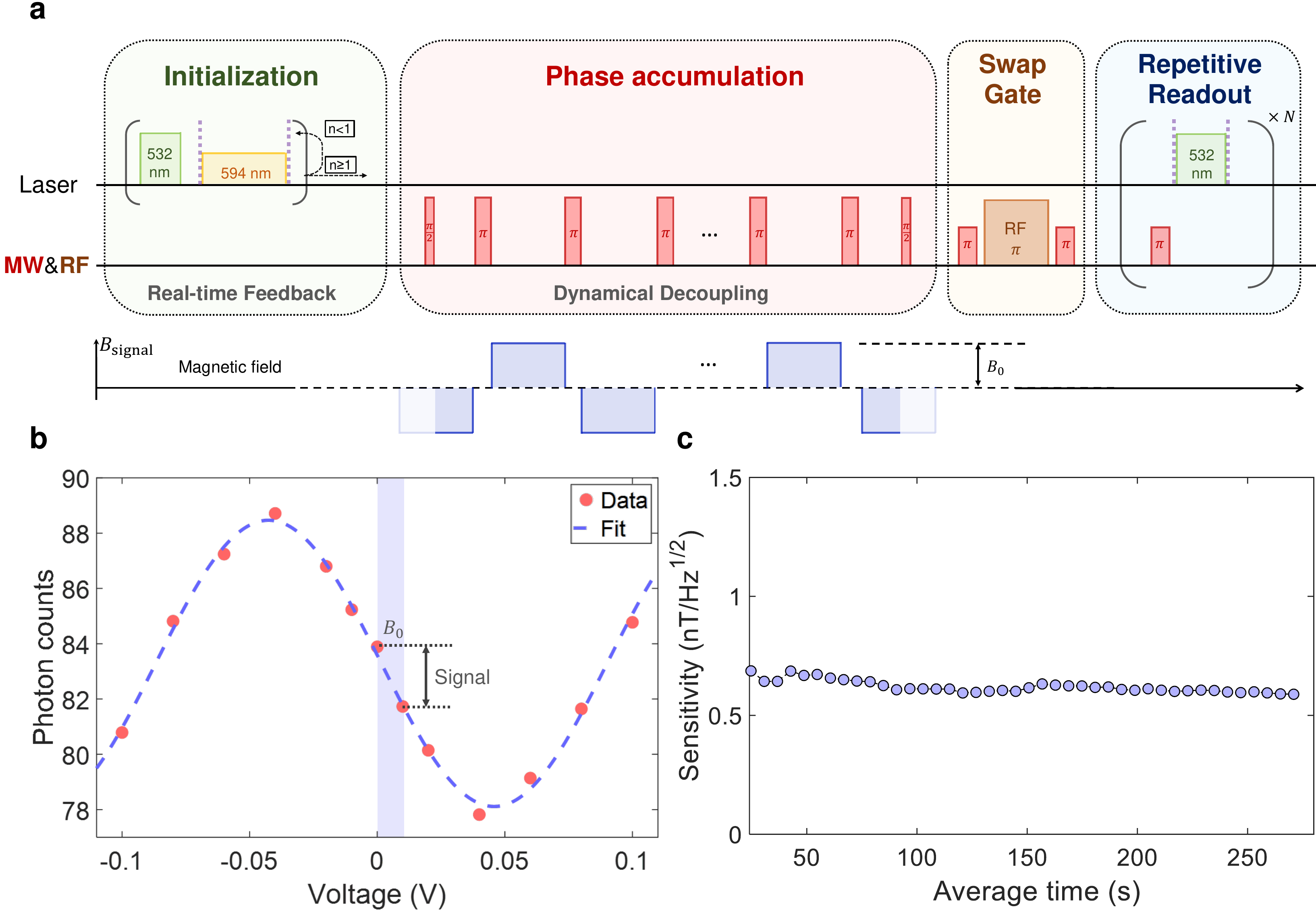}
\caption{\label{sequence}\textbf{Magnetic field measurement.} \textbf{a,} The experimental sequence comprised of initialization, phase accumulation, and readout.
   Real-time feedback is performed to initialize the NV charge state.
   The magnetic field is measured by encoding it into an accumulated phase using dynamical decoupling sequences.
   The technique of repetitive readout is employed to improve the readout fidelity.
   Fluorescence photons are collected during the intervals between two purple dashed lines.
   The waveform $B_{\rm signal}$ is generated by the AWG and fed into a coil to produce a small magnetic field. See more details in Fig. S4 of the Supplemental Material.
 \textbf{b,} The interference pattern for magnetic sensing as a function of the AWG voltage. The dashed line is given by fitting the experimental data. 
\textbf{c,} The magnetic sensitivity measured as a function of the time of data accumulation in the experiment. The beginning deviation from the final value is due to the small data amount, and the values converge as the data accumulates continuously.
}
\end{figure*}

\begin{figure*}\center
\includegraphics[width=0.95\textwidth]{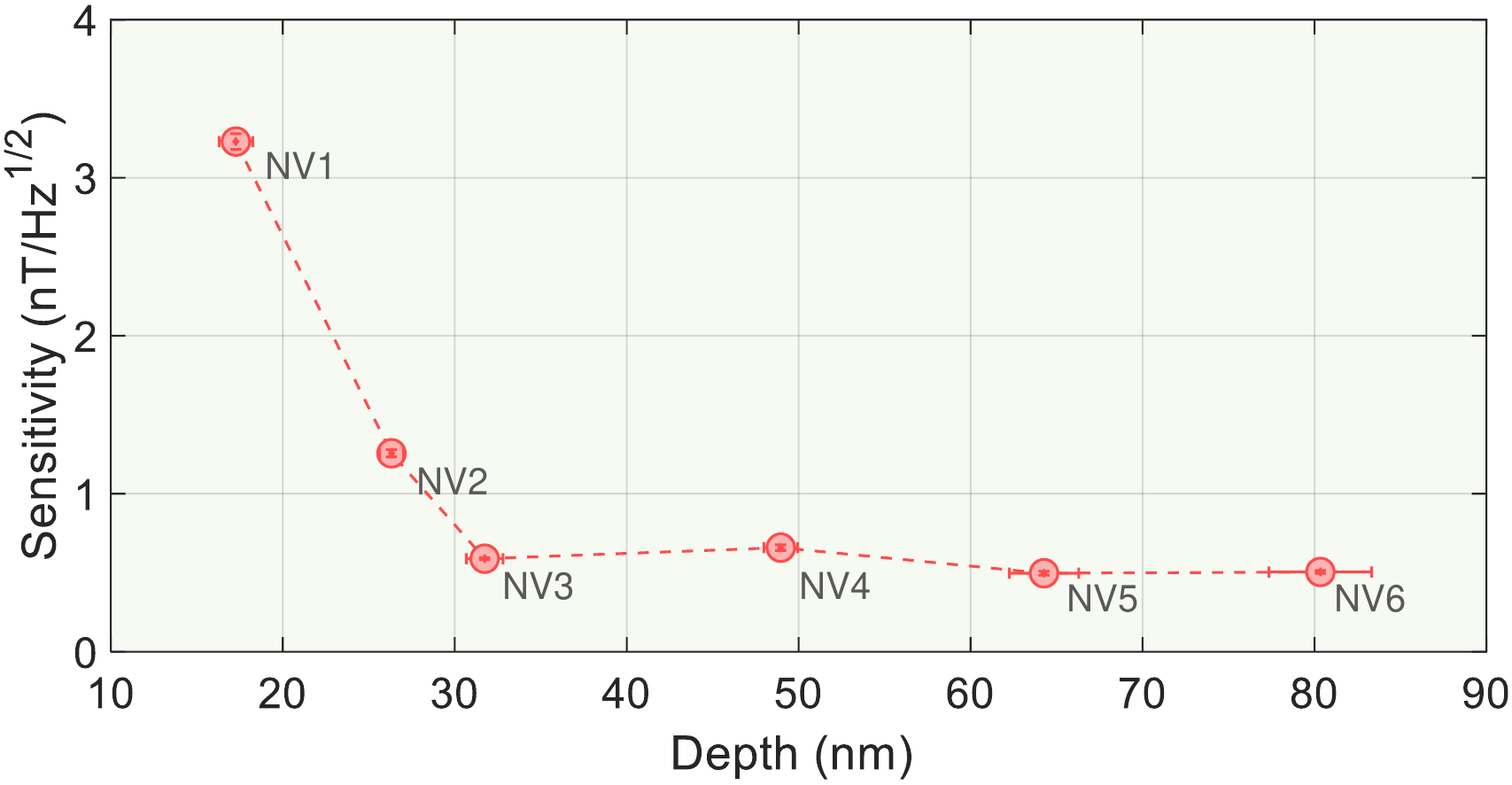}
\caption{\label{result}\textbf{Magnetic field sensitivity.}  Measured magnetic field sensitivities for six near-surface NV centers with different depths below the diamond surface.
}
\end{figure*}

The interference sequence for magnetic field measurement is plotted in Fig. \ref{sequence}a, composed of initialization, phase accumulation and readout.
The dynamical decoupling sequences in the second block are used to encode the detected magnetic field into the phase of the NV spin.
It is worth noting that the use of high-frequency microwaves (MWs) and the existence of the $^{15}$N hyperfine coupling make it rather difficult to apply strong $\pi/2$ and $\pi$ pulses for the NV electron spin.
Therefore, shaped $\pi/2$ and $\pi$ pulses optimized by the gradient ascent pulse engineering (GRAPE) algorithm \cite{Khaneja2005JMR} are adopted here for a lower amplitude and a higher fidelity (see Materials and Methods for more details).
The detected magnetic field is produced by a copper coil into which a waveform output from an arbitrary waveform generator (AWG) is fed.
Changing the AWG voltage varies the number of collected photons in Fig. \ref{sequence}b.
The detected magnetic fields are determined with the coefficient $\approx$112 nT/V, which is obtained by fitting the interference pattern.
With a specific magnetic field applied, the sensitivity of our system is given by the asymptotic behavior in Fig. \ref{sequence}c, after averaging the data for several hundred seconds (see Materials and Methods).
To find how the sensitivity is affected by the noise from the diamond surface, NV centers with different depths are investigated. 
By repeating the established procedures above, the sensitivities for six NV centers are measured and displayed in Fig. \ref{result}. 
The sensitivity is improved gradually as the NV depth increases, and saturates at about 0.5 $\rm nT/\sqrt{Hz}$.
It implies that the effect of the surface noise is no longer the major limiting factor when the depth reaches $\sim$30 nm.
By the way, the sensitivity of 0.5 $\rm nT/\sqrt{Hz}$ is the highest record for a single NV center at room temperature.

\begin{figure*}\center
\includegraphics[width=1.0\textwidth]{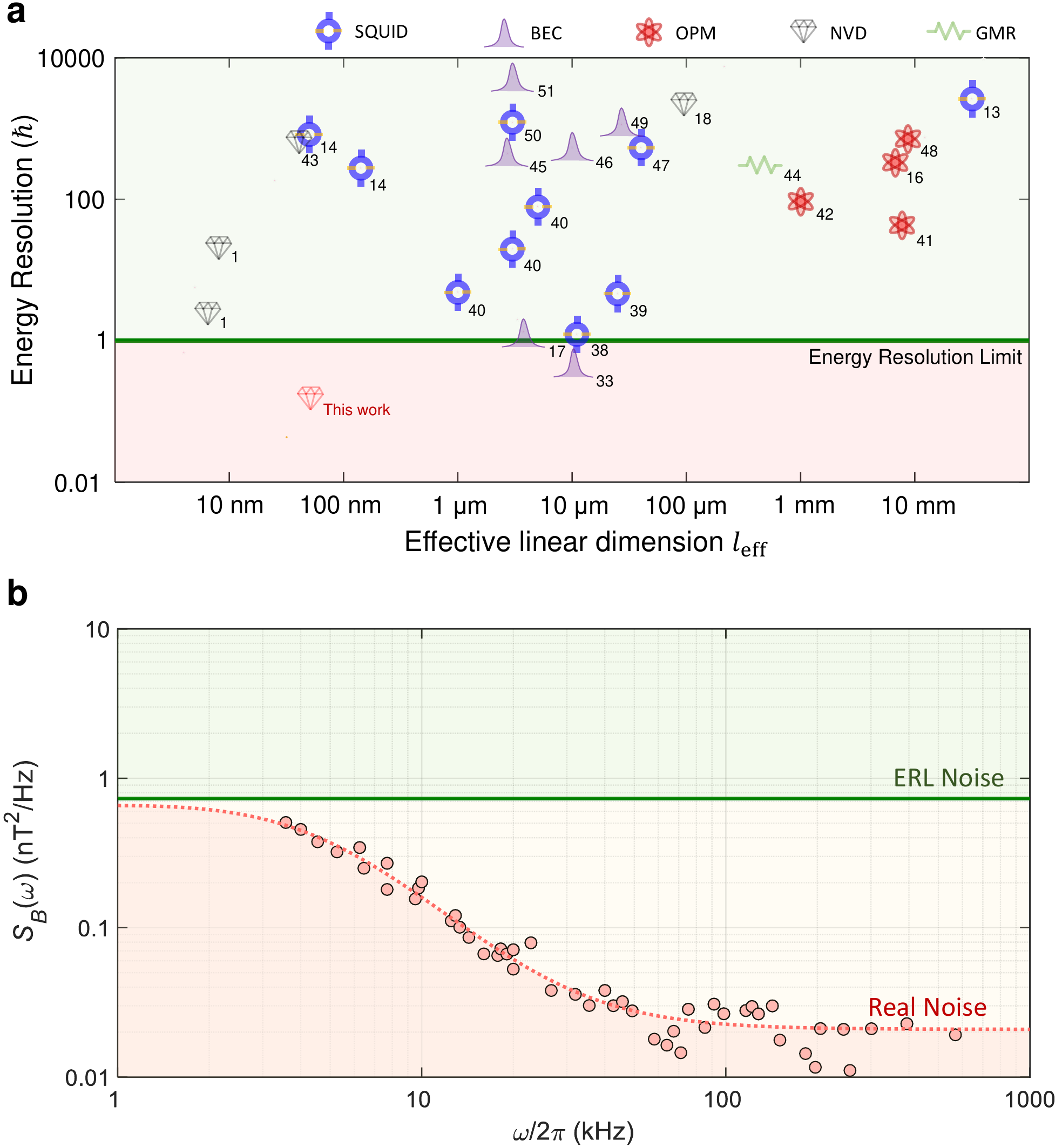}
\caption{\label{noise}\textbf{Energy resolution per bandwidth and measured noise spectrum.} \textbf{a,} Reported energy resolution per bandwidth $E_R$ for different magnetometers versus their effective linear dimensions $l_{\rm eff}$.
   Line shows $E_R\equiv{\langle\delta B^2\rangle Tl^3_{\rm{eff}}}/{(2\mu_0)}=\hbar$.
   The energy resolution of NV3 in Fig. \ref{result} is labeled with ``This work''.
   SQUID, superconducting quantum interference device; BEC, Bose-Einstein condensate; OPM, optically pumped magnetometer; NVD, nitrogen-vacancy center in diamond; GMR, giant magnetoresistance. See more details in References with the labeled numbers and in Table S1 of the Supplemental Material.
 \textbf{b}, The noise spectral density of the same NV in (a) is measured under multiple dynamical decoupling sequences.
   The red dashed line is the Lorentzian fitting.
   The noise level constrained by $E_R=\hbar$ is indicated by the green solid line (ERL noise).
   The noise densities for the frequencies greater than 100 kHz are below the ERL noise.
}
\end{figure*}

{\subsection*{{Energy resolution}}}
{
In the following discussions, the energy resolution limit (ERL) is adopted to benchmark the NV sensors constructed above.
{As discussed in the literature} \cite{Mitchell2020RMP}, the ERL quantitatively depicts the contradiction between the spatial resolution and the sensitivity.
{Among more than twenty kinds of magnetometers, SQUID, BEC, OPM, and NV centers have demonstrated the energy resolution near one $\hbar$ \cite{Mitchell2020RMP}, and all of them have practical importance. Thus, mainly these four types of magnetometers are included in Fig. \ref{noise}a for comparison.}
The energy resolution combines field, temporal, and spatial resolution in a concise form, which simply reads}
\begin{equation}
\label{erl}
E_R\equiv\frac{\langle\delta B^2\rangle T l^3_{\rm{eff}}}{2\mu_0}\geq\hbar
\end{equation}
where $E_R$ is the energy resolution per bandwidth, and greater than the Planck's constant $\hbar$. In the expression of $E_R$, $\mu_0$ is the vacuum permeability, $T$ is the measurement time, $\langle\delta B^2\rangle$ is the magnetic field variance, and the sensitivity $\eta_B=\langle\delta B^2\rangle^{1/2}\sqrt{T}$.
The symbol $l_{\rm{eff}}$ represents the effective linear dimension of the sensor, and determines the spatial resolution.

The ERL is actually not a universally proven quantum limit, though strictly derived for SQUIDs \cite{Tesche1977} and spin-precession ensembles \cite{Mitchell2020} under some general assumptions.
{Recently, a theoretical scheme has been proposed to surpass the ERL by using ferromagnetic torque sensors \cite{Vinante2021PRL}, and an experiment has achieved the energy resolution below $\hbar$ with single-domain Bose–Einstein condensate magnetometer ($E_R=0.48 \hbar$ and $E_R=0.075 \hbar$ if the duty cycle of experiment is not considered)~\cite{Palacios2022PNAS}.
Additionally, single quantum systems (SQSs), e.g. single NV centers and trapped ions, also have the potential to achieve the energy resolution below $\hbar$ \cite{Mitchell2020RMP}.}
It is worth noting that the effective linear dimension $l_{\rm{eff}}$ for SQSs is not the size of the system wavefunction {(ultimately limiting the spatial resolution), but the achieved spatial resolution when SQSs are used to detect external objects such as a piece of material in condensed matter physics\cite{Chibotaru2000Nature} or a protein in structural biology\cite{Lovchinsky2016Science}, not internal objects like residual atoms in vacuum~\cite{Grier2009PRL} or spins in diamond lattice~\cite{Abobeih2019Nature}.}
In this regard, as analogous to ensemble sensors (e.g. OPM, BEC, and NV ensembles), shrinking $l_{\rm{eff}}$ for SQSs always dramatically increases the noise and then shortens coherence times, for which the measurements performed by SQSs are still not below $\hbar$ either \cite{Lovchinsky2016Science, Fang2013PRL, Mitchell2020RMP}.

To find the {best} energy resolution in our work, by multiplying the spatial resolutions achieved by detecting external nuclear spins respectively, the energy resolutions for six NV centers are obtained (see Table S2 of the Supplemental Material).
The energy resolution given by the 31.7-nm-deep NV center (NV3 in Fig. \ref{result}) is {0.18 $\pm$ 0.01 $\hbar$}. {As displayed in Fig. \ref{noise}a, {it gives the best energy resolution among the four types of magnetometers.}}

{{Essentially, $E_R=\hbar$ gives the noise level at a certain dimension that the magnetic sensor with that dimension must comply with.}
Therefore, we proceed to investigate the noise felt by the NV center.}
The noise spectrum \cite{Romach2015PRL} for the NV center with the {best} energy resolution (NV3) is derived from its decoherence behaviors under multiple dynamical decoupling sequences (see Materials and Methods and fig. S8).
{As shown in Fig. \ref{noise}b, the measured noise spectral densities are all well below those constrained by $E_R=\hbar$ for the frequencies above 100 kHz. The best energy resolution (0.18 $\hbar$ in this work) can approach 0.03 $\hbar$ if initialization and readout fidelities as well as duty cycles in the experiments are further improved.}

\section*{{Discussion}}
In conclusion, the main drawbacks for near-surface NV centers, such as short coherence times, low-fidelity initialization, and low-efficiency readout, are all elegantly overcome by harnessing multiple quantum techniques.
The highest sensitivity (0.5 $\rm{nT/\sqrt{Hz}}$) at the nanoscale is experimentally achieved, signifying that our sensor takes the combined advantages of magnetic field, temporal, and spatial resolution over all other magnetometers.
Therefore, it opens up a new exploration region on searching new short-range interactions, microscopic magnetic phenomena in condensed matters, and detection of life activities at the macro-molecule scale, {all demanding both good magnetic sensitivity and high spatial resolution}.
The best energy resolution {0.18 $\pm$ 0.01 $\hbar$} is given by combining the achieved sensitivity and spatial resolution.
In the future, it can be further improved by several orders, through reducing the surface noise with surface treatments \cite{Lovchinsky2016Science} and suppressing the spin-lattice relaxation with cryogenic temperatures \cite{Vinante2012PRL}.

\section*{{Materials and Methods}}

\subsection*{Diamond sample and experimental setup}
\label{experimental_setup}
The targeted NV centers reside in a bulk diamond whose top face is perpendicular to the [100] crystal axis and lateral faces are perpendicular to [110]. 
The top layer with a thickness of several micrometers is grown with an isotopically purified carbon source (99.999\% $^{12}$C) and the nitrogen concentration is less than 5 ppb.
The NV centers were created with a density of 5$\times 10^{-8}$cm$^{-2}$ 15 keV and 22.5 keV $\rm{^{15}N^+}$ ion implantation in two different regions and followed by annealing in vacuum for 4 hours at 1000 ℃. 
The luminescence rates of the NV centers are distributed within 150$\sim$230 kcounts/s.
The diamond is mounted on a typical optically detected magnetic resonance confocal setup, synchronized with a microwave bridge by a multichannel pulse blaster (Spincore, PBESR-PRO-500). The 532-nm green laser and the 594-nm orange laser for driving NV electron dynamics, and sideband fluorescence (650–800 nm) go through the same oil objective (Olympus, UPLSAPO 100XO, NA 1.40). To protect the NV center’s negative state and longitudinal relaxation time against laser leakage effects, all laser beams pass twice through acousto–optic modulators (AOM) (Gooch $\&$ Housego, power leakage ratio $\sim$1/1,000) before they enter the objective. The fluorescence photons are collected by avalanche photodiodes (APD) (Perkin Elmer, SPCM-AQRH-14) with a counter card (National Instruments, 6612). The ZI AWG also has a built-in counter to perform real-time feedback for preparing NV negative state. The 18.6 GHz and 24.3 GHz MW pulses for manipulating the NV three sublevels are generated from the microwave bridge, coupled with 0.1-10 MHz radio-frequency (RF) pulses for the nuclear spins via a diplexer, and fed together into the coplanar waveguide microstructure. The external magnetic field ($\approx$7662 G) is generated from a permanent magnet and aligned parallel to the NV axis through a three-dimensional positioning system. The positioning system, together with the platform holding the diamond and the objective, is placed inside a thermal insulation copper box. The temperature inside the copper box stabilizes down to a sub-mK level through the feedback of the temperature controller (Stanford, PTC10). The overall setup is plotted in Fig. S2.

\subsection*{Contradictions between multiple techniques and solutions}
The techniques used here are contradictory to a certain degree and demand many technical improvements for being combined together. The contradictions between them and the corresponding solutions are graphically represented by Fig. S1 and summarized here as follows:
\begin{enumerate}
\item If the techniques of charge initialization and repetitive readout are arbitrarily used, due to long times consumed by them, they will not improve the sensitivity but in turn lower it by reducing the duty cycle. In this work, the coherence times for near-surface NV centers are extended to several milliseconds comparable to those consumed by them, and these two techniques are elegantly optimized to largely enhance the sensitivity.
\item The high magnetic field used for repetitive readout poses challenges to the coherent manipulation of the NV spin in the experiments. In this work, it is solved by using low-amplitude and high-fidelity shaped pulses.
\item The coherence time is dramatically shortened when the NV center draws near the diamond surface. In this work, the technique of dynamical decoupling is used to suppress the surface noise and prolong the coherence time.
\item Continuously running experiments without reducing the duty cycle is challenging due to the field instability under a high magnetic field ($\sim$ 7662 G in this work). In this work, the high magnetic field is stabilized below 0.1 G through active temperature feedback, without the need to frequently calibrate MW frequency.
\end{enumerate}

\subsection*{Duty cycle}
\label{magnetic_stability}
The duty cycle in our experiments are mainly affected by the initialization and readout time and the time for calibrating MW frequency. With the techniques of real-time feedback and repetitive readout, both the initialization fidelity and the readout fidelity can be improved above 99\%, but it takes a great amount of time and dramatically reduces the experimental duty cycle. The cycle numbers for both processes are therefore optimized to realize the best sensitivity for each NV center. Besides, the high magnetic field used here must be stabilized, otherwise frequently calibrating MW frequency during the experiment will reduce the duty cycle. 

\subsection*{Magnetic field stability}
{In this work, the magnetic field is stabilized below 0.1 G without the need for time-consuming calibration. Temperature fluctuation is the most important factor which affects the magnetic field stability. The temperature disturbs the magnetic field mainly by changing the magnetization of the magnets (0.12$\%$ \textcelsius$^{\text{-1}}$ for NdFeB magnets) and the NV position in the magnetic field. With an elaborately designed configuration for heat isolation and suitable PID parameters for feedback, the temperature fluctuation is suppressed down to less than 1 mK. The latter is more remarkable especially for a field gradient of over 2 G/\textmu m at high magnetic fields (7662 G). Changes in temperature and humidity outside the temperature-controlled box can affect the distance between the sample and the magnet through the optical platform. The Fig. S9 shows the shift of the resonance spectral peaks during the experiments, which shows that the bias field oscillates within $\sim$0.1 G. The oscillation is caused by the periodic cooling of the air conditioner (through the optical table thermal expansion) affecting the distance between the magnet and the sample. As for the temperature control system, the main attention should be paid to heat isolation since PID feedback is easily achieved at the best level (only noise left with no distinct oscillation).  }

\subsection*{The system Hamiltonian}
The NV electron spin ($S=1$) in the ground state of the NV$^-$ triplet and the adjacent $^{15}$N nuclear spin ($I=1$) comprise our nanoscale system. The Hamiltonian with an external magnetic field  $B_0 \approx$ 7662 G applied along the axis of the NV is given by
\begin{equation}
\label{Hamiltonian_0}
H_0 \approx \overbrace{D S_z^2+\gamma _e B_0 S_z}^{\rm{NV}}+\overbrace{\gamma _{\rm N} B_0 I_z^{\rm N}+A_\parallel S_z I_z^{\rm N}}^{^{15}\rm{N}}
\end{equation}
where $\gamma _e$ and $\gamma _{\rm N}$ are the gyromagnetic ratios of the electron spin and the $^{15}$N nuclear spin respectively. $S_z$ and $I_z^{\rm N}$ are the components of three spin operators along the axis of the NV. The hyperfine interaction $A_\parallel$ is roughly 3.03 MHz. $D\approx2870$ MHz is the zero-field splitting of the NV ground state. MW and RF with multiple frequencies are imposed to coherently control the electron spin and the $^{15}$N nuclear spin with the control Hamiltonian
\begin{equation}
\label{Hamiltonian_c}
H_c(t) = \Omega^{\rm MW}(t)\cos(\omega ^{\rm MW} t+\phi^{\rm MW}(t))S_x+\Omega^{\rm RF}(t)\cos(\omega ^{\rm RF} t+\phi^{\rm RF}(t))I_x^{\rm N}
\end{equation}
We implemented the quantum circuit (Fig. 2a) by transforming into a suitable interaction picture with some rotation-wave approximations.

\subsection*{Determination of NV depth}
In our work, the depth of the NV center is obtained by detecting the nuclear magnetic resonance signal from the proton sample put upon the diamond surface.
Glycerine or immersion oil (IMMOIL-F30CC) of the objective is placed upon the diamond surface and the fluctuating magnetic field from the $^1\mathrm{H}$ nuclear spins in the samples is measured using a XY16-N dynamical decoupling pulse sequence. The fluctuating magnetic field causes an extra decoherence of the NV spin, which is described by\cite{Pham2016PRB}
\begin{equation}
\label{decoherence}
C(\tau) \approx \exp \left[-\frac{2}{\pi^{2}} \gamma_{e}^{2} B_{\mathrm{RMS}}^{2} K(N \tau)\right]
\end{equation}
where $\gamma_{e}$ is the electron gyromagnetic ratio and $B_{\mathrm{RMS}}^{2}$ is the mean square magnetic field produced by the proton spins. $K(N \tau)$ depends on the pulse sequence, and the coherence time $T_{2n }^*$ and the diffusion coefficient of the proton spins\cite{Pham2016PRB}.
For a diamond with a [100] top surface, $B_{\mathrm{RMS}}^{2}$ for the NV depth $d_{\mathrm{NV}}$ is
\begin{equation}
\label{brms}
B_{\mathrm{RMS}}^{2}=\rho\left(\frac{\mu_{0} \hbar \gamma_{n}}{4 \pi}\right)^{2}\left(\frac{5 \pi}{96 d_{\mathrm{NV}}^{3}}\right)
\end{equation}
where $\gamma_{n}$ is the gyromagnetic ratio of $^1\mathrm{H}$ spins, and $\rho$ is the nuclear spin number density ($\rho=66 \mathrm{~nm}^{-3}$ for glycerine by calculation, and $\rho=69.5 \mathrm{~nm}^{-3}$ for the immersion oil measured by EDUMR20-015V-I NMR system).
The NV depths can be extracted by fitting with the equations above.
The results are summarized below, and more details can be found in Fig. S3.
\begin{table}[H]\footnotesize
\centering
\setlength{\tabcolsep}{10pt}
\renewcommand{\arraystretch}{1.6}
\begin{tabular}{ccccccc}
\hline
№          & 1          & 2          & 3          & 4          & 5          & 6          \\ \hline
Depth (nm) & 17.3 ± 1.0 & 26.3 ± 0.7 & 31.7 ± 1.1 & 49.0 ± 1.0 & 64.3 ± 2.0 & 80.3 ± 3.0 \\ \hline
\end{tabular}
\end{table}

\subsection*{Measurement of magnetic sensitivity}
In order to determine the magnetic sensitivity experimentally, we first have to measure the number of collected photons $N_{\rm ph}$ against the AWG voltage $V$ proportional to the amplitude of the small magnetic field produced by the coil.
The magnetic field magnitude per unit voltage $B_V$ is obtained by fitting with
\begin{equation}
N_{\rm ph}(v)=a\sin(\gamma_e T B_V V+\phi)+c
\end{equation}
where $T$ is the interrogation time, $\gamma_e$ is the gyromagnetic ratio of the NV spin. $B_V$, $a$, and $c$ are the fitting parameters.
Subsequently, we apply a specific magnetic signal and measure the signal-to-noise ratio (SNR) over a range of experimental times. 
The sensitivity is then determined by
\begin{equation}
\label{snoise1}
\text{Sensitivity} = \frac{\text{Signal amplitude}}{\text{SNR per } \sqrt{\text{unit time}}}
\end{equation}
The determined sensitivity is approximately equal to that estimated based on Eq. \ref{sensitivity} {(see Table S2)}.
The optimized parameters for six centers are summarized in Table S2, and the corresponding results are displayed in Fig. 2 and Fig. S5.

{The method for estimating the sensitivity based on Eq. \ref{sensitivity} is given here.
The fidelity of conventional readout is $F_{R_0}=[{{1+2(\alpha_0+\alpha_1)/(\alpha_0-\alpha_1)^2}}]^{-1/2}$,
where $\alpha_0$ and $\alpha_1$ are the average numbers of the collected photons by reading out the states $|0\rangle$ and $|\pm1\rangle$ for one time.
As for repetitive readout, the fidelity is given by $F_{R_N}=[{{1+2(\alpha_0+\alpha_1)/(N(\alpha_0-\alpha_1)^2)}}]^{-1/2}=[{{1+(1/F_{R_0}^2-1)/N}}]^{-1/2}$, where $N$ is the number of the readout cycles. 
The initialization fidelity of the NV$^{-}$ charge state is approximately 92\%. 
The population $p$ of the electron state $|0\rangle$ is approximately 90\%, which affects the final sensitivity by a factor of $(3p-1)/2$. The interrogation time for sensing a small magnetic signal is determined as the length of the dynamical decoupling sequence after which the coherence remains roughly 0.5$-$0.6.
The relevant experimental parameters of each NV center are gathered in Table S2. Based on Eq. \ref{sensitivity}, the sensitivity of each NV center is estimated with the results listed in Table S2.
Overall, the estimation results are consistent with the measured sensitivities. 
Small deviations may originate, on the one hand, from imperfect implementations of higher-order dynamical decoupling sequences due to MW amplitude fluctuations, and on the other hand, from the difference between the actual coherence (0.5$-$0.6) and that used for the estimation (fixed as 0.6).}

{Besides, it is worth noting that the parameters of the dynamical decoupling sequence used by each NV center just give a lower limit to the measurement bandwidth, and the same sensitivity can be achieved by increasing the order of the decoupling sequence for sensing magnetic signals with higher frequencies. These NV centers have a common bandwidth at higher frequencies, and thus, although the parameters used for each NV center are different, the results can be put together for comparison.}

\subsection*{Noise analysis}
The analysis of the noise felt by the NV center is based on the technique of spectral decomposition\cite{Romach2015PRL}.
Generally, the spin coherence decays as a function of time
\begin{equation}
\label{noise1}
C(t) = \exp({-\Delta\phi^2(t)/2})
\end{equation}
due to the interactions with surrounding spins. 
The phase variance $\Delta\phi^2(t)$ depends on the noise spectrum $S(\omega)$ of the environment and the filter function $F_{T}(\omega)$ of the dynamical decoupling sequence
\begin{equation}
\label{noise2}
\Delta\phi^2(t)=\gamma_e^2 \int_{0}^{\infty} S(\omega) {F_{T}(\omega)}\frac{d \omega}{\pi}
\end{equation}
The filter function for high-order dynamical decoupling sequences can be approximated as:
\begin{equation}
\label{noise3}
F_{T}(\omega)\approx2\pi T\cdot\frac{4}{\pi^2}\sum_{k=-\infty}^{\infty}\frac{\delta(\omega-(2k+1)\omega_0)}{(2k+1)^2}
\end{equation}
where $T$ is the total evolution time and $\omega_0=\pi N/T$.
Combining Eq. \ref{noise1}, \ref{noise2} and \ref{noise3}, the noise spectral density at frequency $\omega_0$ is given by
\begin{equation}
\label{noise4}
\frac{8}{\pi^2}\sum_{k=0}^{\infty}\frac{S((2k+1)\omega_0)}{(2k+1)^2}=\frac{-2\ln C(T)}{\gamma_e^2T}
\end{equation}
In order to acquire a wide-range noise spectrum, multiple dynamical decoupling sequences with different orders are executed, as shown in Fig. S8.
The effect of spin-lattice relaxation is deducted from the decay curve for millisecond-scale spin coherence.
Besides, according to Eq. \ref{noise4}, we build an iterative formula for a more accurate calculation
\begin{gather}
\label{noise5}
S_{n}(\omega_0)=\frac{\pi^2}{8}S_0(\omega_0)-\sum_{k=1}^{\infty}\frac{S_{n-1}((2k+1)\omega_0)}{(2k+1)^2}
\end{gather}
with $S_0(\omega_0)$ the right side of Eq. \ref{noise4}.
When the iteration number $n$ is large enough, $S_n(\omega_0)$ converges to the noise spectrum $S(\omega_0)$, and generally, n=1 or 2 (n=1 in our case) is enough due to its intrinsic fast convergence.
Besides, the noise level constrained by the ERL in Fig. 4 is given by
\begin{equation}
S(\omega)=\frac{2\mu_0\hbar}{e\cdot l_{\text{eff}}^3},
\end{equation}
where the number $e$ in the denominator is originated from the spin decoherence during phase accumulation.

\subsection*{Real-time feedback for NV$^-$ preparation}
Since the zero-phonon line for NV$^-$ is 637 nm and for NV$^0$ is 575 nm, the 594-nm orange laser can excite the NV$^-$ but can not excite NV$^0$.
Therefore, the negative charge state is prepared if one photon is collected during laser illumination\cite{Xie2021SciAd,{charge_2020}}.
If not, repeat it until one photon is collected. 
Before photon counting during each cycle, the green laser is switched on to mix charge states. 
For evading the reduction of duty cycle, the experimental parameters are carefully optimized: 970 ns for state readout, 95 ns for state mixing, and 100 cycles for the upper bound of feedback loops.
The results are shown in Fig. S7.

\subsection*{Optimal control}
The high magnetic field ($\approx$7662 G) poses challenges to the coherent manipulation of the NV spin in the experiment. 
The existence of the $^{15}$N hyperfine coupling demands rather strong MW pulses for simultaneous manipulation of the NV spin in two subspaces of the $^{15}$N spin states.
However, the use of high-frequency MWs in high magnetic field, due to the great attenuation loss in the circuits, makes it difficult to achieve such strong MW pulses.
To achieve high-fidelity manipulation with fairly weak MW pulses, the technique of optimal control is employed in dynamical decoupling sequences, and the shaped pulses are optimized numerically by the GRAPE (gradient ascent pulse engineering) algorithm detailed in Fig. S6.

\section*{Signal of a proton}
{The magnetic signal of a single proton located right above the NV with a depth of $d_{\mathrm{NV}}$ is given by \begin{equation}
\label{noise4}
\eta_p=\frac{\mu_0\hbar\gamma_p}{4\pi}\cdot\frac{1}{{d_{\mathrm{NV}}}^3}
\end{equation}
where $\gamma_p$ is the gyromagnetic ratio of the proton spin, and $\mu_0$ is the vacuum permeability. The sensitivity needed for detecting a single proton (signal-to-noise ratio (SNR) of 1) with 1 s (100 s) of data accumulation is calculated as a function of NV depth, which is plotted as a blue solid (green dashed) line in the Fig. S10. The minimum number of detectable proton nuclear spins is 6.7 for the NV with a depth of 31.7 nm (data accumulation for 1 s, SNR = 1).}

\bibliographystyle{Science}

\begin{thebibliography}{40}

\bibitem{Lovchinsky2016Science} Lovchinsky I, Sushkov AO and Urbach E \textit{et al.} Nuclear magnetic resonance detection and spectroscopy of single proteins using quantum logic. \textit{Science} \textbf{351}, 836-841 (2016).

\bibitem{Safronova2018RMP} Safronova MS, Budker D and DeMille D \textit{et al.} Search for new physics with atoms and molecules. \textit{Rev. Mod. Phys.} \textbf{90}, 025008 (2018).

\bibitem{Rong2018PRL} Rong X, Jiao M and Geng J \textit{et al.} Constraints on a spin-dependent exotic interaction between electrons with single electron spin quantum sensors. \textit{Phys. Rev. Lett.} \textbf{121}, 080402 (2018).

\bibitem{Sikivie2021RMP} Sikivie P. Invisible axion search methods. \textit{Rev. Mod. Phys.} \textbf{93}, 015004 (2021).

\bibitem{Blatter1994RMP} Blatter G, Feigel'man MV and Geshkenbein VB \textit{et al.} Vortices in high-temperature superconductors. \textit{Rev. Mod. Phys.} \textbf{66}, 1125 (1994).

\bibitem{Cheong2020npj} Cheong SW, Fiebig M and Wu W \textit{et al.} Seeing is believing: visualization of antiferromagnetic domains. \textit{npj Quantum Mater.} \textbf{5}, 3 (2020).

\bibitem{Chibotaru2000Nature} Chibotaru LF, Ceulemans A and Bruyndoncx V. Symmetry-induced formation of antivortices in mesoscopic superconductors. \textit{Nature} \textbf{408}, 833-835 (2000).

\bibitem{Wishart1991jmb} Wishart DS, Sykes BD and Richards FM. Relationship between nuclear magnetic resonance chemical shift and protein secondary structure. \textit{J. Mol. Biol.} \textbf{222}, 311-333 (1991).

\bibitem{Glover2002RPP} Glover P and Mansfield P. Limits to magnetic resonance microscopy. \textit{Rep. Prog. Phys.} \textbf{65}, 1489 (2002).

\bibitem{Frederick2015Cell} Frederick KK, Michaelis VK and Corzilius B \textit{et al.} Sensitivity-enhanced NMR reveals alterations in protein structure by cellular milieus. \textit{Cell} \textbf{163}, 620-628 (2015).

\bibitem{Abobeih2019Nature} Abobeih MH, Randall J and Bradley CE \textit{et al.} Atomic-scale imaging of a 27-nuclear-spin cluster using a quantum sensor. \textit{Nature} \textbf{576}, 411-415 (2019).

\bibitem{Vinante2021PRL} Vinante A, Timberlake C and Budker D \textit{et al.} Surpassing the Energy Resolution Limit with ferromagnetic torque sensors. \textit{Phys. Rev. Lett.} \textbf{127}, 070801 (2021).

\bibitem{Schmelz2016IEEE} Schmelz M, Zakosarenko V and Chwala A \textit{et al.} Thin-Film-Based Ultralow Noise SQUID Magnetometer. \textit{IEEE Trans. Appl. Supercond.} \textbf{26}, 1 (2016).

\bibitem{Vasyukov2013NatureNano} Vasyukov D, Anahory Y and Embon L \emph{et al.} A scanning superconducting quantum interference device with single electron spin sensitivity. \emph{Nat. Nanotechnol.} \textbf{8}, 639-644 (2013).

\bibitem{Robbes2006Sensor} Robbes D. Highly sensitive magnetometers—a review. \emph{Sens. Actuator A Phys.} \textbf{129}, 86-93 (2006).

\bibitem{Kominis2003Nature} Kominis IK, Kornack TW and Allred JC \textit{et al.} A subfemtotesla multichannel atomic magnetometer. \emph{Nature} \textbf{422}, 596-599 (2003).

\bibitem{Vengalattore2007PRL} Vengalattore M, Higbie JM and Leslie SR \emph{et al.} High-Resolution Magnetometry with a Spinor Bose-Einstein Condensate. \emph{Phys. Rev. Lett.} \textbf{98}, 200801 (2007).

\bibitem{Wolf2015PRX} Wolf T, Neumann P and Nakamura K \emph{et al.} Subpicotesla Diamond Magnetometry. \emph{Phys. Rev. X} \textbf{5}, 041001 (2015).

\bibitem{Mitchell2020RMP} Mitchell MW and Alvarez SP. \emph{Colloquium}: Quantum limits to the energy resolution of magnetic field sensors. \emph{Rev. Mod. Phys.} \textbf{92}, 021001 (2020).

\bibitem{Staudacher2013Science} Staudacher T, Shi F and Pezzagna S \emph{et al.} Nuclear magnetic resonance spectroscopy on a (5-nanometer)$^3$ sample volume. \emph{Science} \textbf{339}, 561-563 (2013).

\bibitem{Pham2016PRB} Pham LM, DeVience SJ and Casola F \emph{et al.} NMR technique for determining the depth of shallow nitrogen-vacancy centers in diamond. \emph{Phys. Rev. B} \textbf{93}, 045425 (2016).

\bibitem{Degen2012PRB} Ofori-Okai BK, Pezzagna S and Chang K \emph{et al.} Spin properties of very shallow nitrogen vacancy defects in diamond. \emph{Phys. Rev. B} \textbf{86}, 081406(R) (2012).

\bibitem{Herbschleb2019NC} Herbschleb ED, Kato H and Maruyama Y \emph{et al.} Ultra-long coherence times amongst room-temperature solid-state spins. \emph{Nat. Commun.} \textbf{10}, 3766 (2019).

\bibitem{Xie2021SciAd} Xie T, Zhao Z and Kong X \emph{et al.} Beating the standard quantum limit under ambient conditions with solid-state spins. \emph{ Sci. Adv.} \textbf{7}, eabg9204 (2021).

\bibitem{Bluvstein2019PRL} Bluvstein D, Zhang Z, Jayich ACB. Identifying and mitigating charge instabilities in shallow diamond nitrogen-vacancy centers. \emph{Phys. Rev. Lett.} \textbf{122}, 076101 (2019).

\bibitem{Du2009Nature} Du J, Rong X and Zhao N \emph{et al.} Preserving electron spin coherence in solids by optimal dynamical decoupling. \emph{Nature} \textbf{461},1265-1268 (2009).

\bibitem{charge_2020} Hopper DA, Lauigan JD and Huang TY \emph{et al.} Real-Time Charge Initialization of Diamond Nitrogen-Vacancy Centers for Enhanced Spin Readout. \textit{Phys. Rev. Appl.} \textbf{13}, 024016 (2020).
%
\bibitem{Liang2009Science} Jiang L, Hodges JS and Maze JR \emph{et al.} Repetitive readout of a single electronic spin via quantum logic with nuclear spin ancillae. \emph{Science} \textbf{326}, 267-272 (2009).

\bibitem{Neumann2010Science} Neumann P, Beck J and Steiner M \emph{et al.} Single-shot readout of a single nuclear spin. \emph{Science} \textbf{329}, 542-544 (2010).
%


\bibitem{Khaneja2005JMR} Khaneja N, Reiss T and Kehlet C \emph{et al.} Optimal control of coupled spin dynamics: design of NMR pulse sequences by gradient ascent algorithms. \emph{J. Magn. Reson.} \textbf{172}, 296-305 (2005).

\bibitem{Tesche1977} Tesche C D and Clarke J. dc SQUID: Noise and optimization. \emph{J. Low Temp. Phys} \textbf{29}, 301 (1977).

\bibitem{Mitchell2020} Mitchell MW. Scale-invariant spin dynamics and the quantum limits of field sensing. \emph{New J. Phys.} \textbf{22}, 053041 (2020).

\bibitem{Palacios2022PNAS} Palacios Alvarez S, Gomez P and Coop S \emph{et al.} Single-domain Bose condensate magnetometer achieves energy resolution per bandwidth below $\hbar$. \emph{Proc. Natl. Acad. Sci.} \textbf{119}, e2115339119 (2022).

\bibitem{Grier2009PRL} Grier AT, Cetina M and Oručević F \emph{et al.} Observation of cold collisions between trapped ions and trapped atoms. \emph{Phys. Rev. Lett.} \textbf{102}, 223201 (2009).

\bibitem{Fang2013PRL} Fang K, Acosta VM and Santori C \emph{et al.} High-Sensitivity Magnetometry Based on Quantum Beats in Diamond Nitrogen-Vacancy Centers. \emph{Phys. Rev. Lett.} \textbf{110}, 130802 (2013).

\bibitem{Romach2015PRL} Romach Y, Müller C and Unden T \emph{et al.} Spectroscopy of surface-induced noise using shallow spins in diamond. \emph{Phys. Rev. Lett.} \textbf{114}, 017601 (2015).

\bibitem{Vinante2012PRL} Jarmola A, Acosta VM and Jensen K \emph{et al.} Temperature-and magnetic-field-dependent longitudinal spin relaxation in nitrogen-vacancy ensembles in diamond. \emph{Phys. Rev. Lett.} \textbf{108}, 197601 (2012).

\bibitem{Cromar1981APL} Cromar MW and Carelli P. Low-noise tunnel junction dc SQUID's. \emph{Appl. Phys. Lett.} \textbf{38}, 723-725 (1981).

\bibitem{Awschalom1988APL} Awschalom DD, Rozen JR and Ketchen MB \emph{et al.} Low-noise modular microsusceptometer using nearly quantum limited dc SQUIDs. \emph{Appl. Phys. Lett.} \textbf{53}, 2108-2110 (1988).

\bibitem{Schmelz2016SCT} Schmelz M, Zakosarenko V, Schönau T \emph{et al.} Nearly quantum limited nanoSQUIDs based on cross-type Nb/AlOx/Nb junctions. \emph{Supercond. Sci. Technol.} \textbf{30}, 014001 (2017).

\bibitem{Dang2010APL} Dang HB, Maloof AC and Romalis MV. Ultrahigh sensitivity magnetic field and magnetization measurements with an atomic magnetometer. \emph{Appl. Phys. Lett.} \textbf{97}, 151110 (2010).

\bibitem{Griffith2010OE} Griffith WC, Knappe S and Kitching J. Femtotesla atomic magnetometry in a microfabricated vapor cell. \emph{Opt. Express} \textbf{18}, 27167-27172 (2010).

\bibitem{Maletinsky2012NatureNano} Maletinsky P, Hong S and Grinolds MS \emph{et al.} A robust scanning diamond sensor for nanoscale imaging with single nitrogen-vacancy centres. \emph{Nat. Nanotechnol.} \textbf{7}, 320 (2012).

\bibitem{Pannetier2004Science} Pannetier M, Fermon C and Le Goff G \emph{et al.} Femtotesla magnetic field measurement with magnetoresistive sensors. \emph{Science} \textbf{304}, 1648-1650 (2004).

\bibitem{Caspar2013PRL} Ockeloen CF, Schmied R and Riedel MF \emph{et al.} Quantum metrology with a scanning probe atom interferometer. \emph{Phys. Rev. Lett.} \textbf{111}, 143001 (2013).

\bibitem{Eto2013PRA} Eto Y, Ikeda H and Suzuki H \emph{et al.} Spin-echo-based magnetometry with spinor Bose-Einstein condensates. \emph{Phys. Rev. A} \textbf{88}, 031602 (2013).

\bibitem{Fong2005ESI} Fong LE, Holzer JR and McBride KK \emph{et al.} High-resolution room-temperature sample scanning superconducting quantum interference device microscope configurable for geological and biomagnetic applications. \emph{Rev. Sci. Instrum.} \textbf{76}, 053703 (2005).

\bibitem{Sheng2013PRL} Sheng D, Li S and Dural N. Subfemtotesla scalar atomic magnetometry using multipass cells. \emph{Phys. Rev. Lett.} \textbf{110}, 160802 (2013).

\bibitem{Wood2015PRA} Wood AA, Bennie LM and Duong A \emph{et al.} Magnetic tensor gradiometry using Ramsey interferometry of spinor condensates. \emph{Phys. Rev. A} \textbf{92}, 053604 (2015).

\bibitem{Gallop2002PC} Gallop J, Josephs-Franks PW and Davies J \emph{et al.} Miniature dc SQUID devices for the detection of single atomic spin-flips. \emph{Physica C Supercond.} \textbf{368}, 109-113 (2002).

\bibitem{Wildermuth2005Nature} Wildermuth S, Hofferberth S and Lesanovsky I \emph{et al.} Microscopic magnetic-field imaging. \emph{Nature} \textbf{435}, 440-440 (2005).



\end{thebibliography}

\clearpage

\section*{Acknowledgements}
This work was partially carried out at the USTC Center for Micro and Nanoscale Research and Fabrication. This work was supported by the National Key R\&D Program of China (Grant No. 2018YFA0306600, 2016YFA0502400), the National Natural Science Foundation of China (Grant No. 81788101, T2125011, 12274396, 91636217), Innovation Program for Quantum Science and Technology (Grant No. 2021ZD0303204, 2021ZD0302200), the CAS (Grant No. XDC07000000, GJJSTD20200001, QYZDY-SSW-SLH004, YSBR-068), the Anhui Initiative in Quantum Information Technologies (Grant No. AHY050000), the Fundamental Research Funds for the Central Universities, the China Postdoctoral Science Foundation (Grant No. 2021M703110) and the Hefei Comprehensive National Science Center.

\clearpage

\end{document}


\title{Supplemental Material for \\ Sub-nanotesla Sensitivity at the Nanoscale with a Single Spin}
\maketitle
%
\pagebreak[4]

\pagebreak[4]
\begin{table}[H]\scriptsize
\centering
\setlength{\tabcolsep}{15.0pt}
\renewcommand{\arraystretch}{0.91}
\begin{tablenotes}
\item[1]\small {Table S1.} \textbf{A summary of energy resolution per bandwidth for different magnetometers.} For single NV centers, the relation between the effective linear dimension $l_{\text{eff}}$ and the NV depth $d_\text{NV}$ is given by $l^3_{\text{eff}}=\frac{4\pi}{3}\cdot d_\text{NV}^3$.\footnotemark[1] 
\item[1]\tiny 
\end{tablenotes}
\begin{tabular}{ccccccccc}
\hline\hline
\rowcolor[HTML]{D9DBF8} 
\textnumero & Type  & $l_{\text{eff}}$ (m) & $\eta_B$ (T/$\sqrt{\text{Hz}}$) & Ref.  & $E_R (\hbar)$ \\ \hline
1     & NV                           & $ 6.4\times10^{-09} $   & $ 5.3\times10^{-08} $        & 1           & 2.84     \\\rowcolor[HTML]{F9FBFF} 
2     & BEC                             & $ 1.1\times10^{-05} $   & $ 5.0\times10^{-13} $        & 17           & 1.24     \\
3     & SQUID                           & $ 3.7\times10^{-06} $   & $ 2.6\times10^{-12} $        & 38           & 1.34     \\\rowcolor[HTML]{F9FBFF} 
4     & SQUID                           & $ 2.5\times10^{-05} $   & $ 2.8\times10^{-13} $        & 39           & 4.68     \\
5     & SQUID                           & $ 1.0\times10^{-06} $   & $ 3.6\times10^{-11} $        & 40           & 4.89     \\\rowcolor[HTML]{F9FBFF} 
6     & NV                           & $ 8.1\times10^{-09} $   & $ 1.1\times10^{-07} $        & 1           & 23.9     \\
7     & SQUID                           & $ 3.0\times10^{-06} $   & $ 1.4\times10^{-11} $        & 40           & 20.0    \\\rowcolor[HTML]{F9FBFF} 
8     & OPM                    & $ 7.6\times10^{-03} $   & $ 1.6\times10^{-16} $        & 41            & 43.5    \\
9     & SQUID                          & $ 5.0\times10^{-06} $   & $ 1.3\times10^{-11} $        & 40           & 79.7     \\\rowcolor[HTML]{F9FBFF} 
10    & OPM                    & $ 1.0\times10^{-03} $   & $ 5.0\times10^{-15} $        & 42           & 94.3    \\
11    & NV                           & $ 4.0\times10^{-08} $   & $ 5.6\times10^{-08} $        & 43           & 774   \\\rowcolor[HTML]{F9FBFF} 
12    & SQUID                          & $ 1.4\times10^{-07} $   & $ 5.1\times10^{-09} $        & 14           & 278   \\
13    & GMR                           & $ 4.3\times10^{-04} $   & $ 3.2\times10^{-14} $        & 44           & 320   \\\rowcolor[HTML]{F9FBFF} 
14    & OPM                    & $ 6.6\times10^{-03} $   & $ 5.4\times10^{-16} $        & 16            & 330   \\
15    & BEC                   & $ 2.7\times10^{-06} $   & $ 7.7\times10^{-11} $        & 45           & 447   \\\rowcolor[HTML]{F9FBFF} 
16    & BEC                             & $ 1.0\times10^{-05} $   & $ 1.2\times10^{-11} $        & 46           & 543   \\
17    & SQUID                          & $ 4.0\times10^{-05} $   & $ 1.5\times10^{-12} $        & 47           & 543   \\\rowcolor[HTML]{F9FBFF} 
18    & OPM                    & $ 8.7\times10^{-03} $   & $ 5.4\times10^{-16} $        & 48            & 726   \\
19    & SQUID                          & $ 5.0\times10^{-08} $   & $ 4.2\times10^{-08} $        & 14           & 832   \\\rowcolor[HTML]{F9FBFF} 
20    & BEC                    & $ 2.7\times10^{-05} $   & $ 3.9\times10^{-12} $        & 49           & 1,148 \\
21    & SQUID                           & $ 3.0\times10^{-06} $   & $ 1.1\times10^{-10} $        & 50           & 1,233 \\\rowcolor[HTML]{F9FBFF} 
22    & NV                  & $ 9.4\times10^{-05} $   & $ 9.0\times10^{-13} $        & 18           & 2,598 \\
23    & SQUID                          & $ 3.1\times10^{-02} $   & $ 1.5\times10^{-16} $        & 13            & 2,644 \\\rowcolor[HTML]{F9FBFF} 
24    & BEC                  & $ 3.0\times10^{-06} $   & $ 2.3\times10^{-10} $        & 51           & 5,389 \\
25    & BEC                  & $ 1.0\times10^{-05} $   & $ 3.4\times10^{-13} $        & 33 & 0.48 \\
 \hline\hline
\rowcolor[HTML]{F9FBFF} 
\end{tabular}
\end{table}

\footnotetext[1]{Supplementary Reference 1: Mitchell MW and Alvarez SP. \emph{Colloquium}: Quantum limits to the energy resolution of magnetic field sensors. \emph{Rev. Mod. Phys.} \textbf{92}, 021001 (2020).}

\begin{table}[H]\scriptsize
\centering
\setlength{\tabcolsep}{3.75pt}
\renewcommand{\arraystretch}{1.3}
\begin{tablenotes}
\item[1]\small {Table S2.} \textbf{Experimental parameters and results of our system.} ``Total time'' means the total time of a single experiment, including initialiation time, interrogation time, and readout time. $\eta_t$ is the sensitivity calculated based on Eq.1 and $\eta_B$ is the measured sensitivity. 
\item[2]\tiny
\end{tablenotes}
\begin{tabular}{cccccccccc}
\hline\hline
\rowcolor[HTML]{FFF2F1} 
\textnumero & Sequence & Interrogation time & Readout cycles & Total time & Duty cycle & Depth (nm)  & $\eta_t$ (nT/$\sqrt{\rm{Hz}}$)  & $\eta_B$ (nT/$\sqrt{\rm{Hz}}$) & $E_R (\hbar)$ \\ \hline
1  & XY16-128 & 0.15 ms            & 904            & 0.928 ms   & 16.2\% & 17.3 $\pm$ 1.0 &3.2& 3.23 $\pm$ 0.05       & 0.85                   \\
\rowcolor[HTML]{F9FBFF} 
2  & XY16-512 & 0.70 ms            & 1422           & 1.889 ms   &37.1\% & 26.3 $\pm$ 0.7 &0.9& 1.26 $\pm$ 0.02       & 0.46                  \\
3  & XY16-512 & 1.80 ms            & 2025           & 3.336 ms   & 54.0\%& 31.7 $\pm$ 1.1 &0.44& 0.59 $\pm$ 0.01       & 0.18                 \\
\rowcolor[HTML]{F9FBFF} 
4  & XY16-512 & 1.80 ms            & 1870           & 3.247 ms   & 55.4\%& 49.0 $\pm$ 1.0 &0.44& 0.66 $\pm$ 0.02       & 0.81                  \\
5  & XY16-512 & 2.20 ms            & 2030           & 3.586 ms   & 61.3\%& 64.3 $\pm$ 2.0 &0.38& 0.50 $\pm$ 0.01       & 1.1                  \\
\rowcolor[HTML]{F9FBFF} 
6  & XY8-32   & 1.50 ms            & 1560           & 2.623 ms   &57.2\% & 80.3 $\pm$ 3.0 &0.49& 0.51 $\pm$ 0.01       & 2.1                   \\ \hline\hline
   &          &                    &                &            &       &                &&                       &                      
\end{tabular}
\end{table}
\pagebreak[4]

\begin{figure}
\centering
\includegraphics[width=1.0\textwidth]{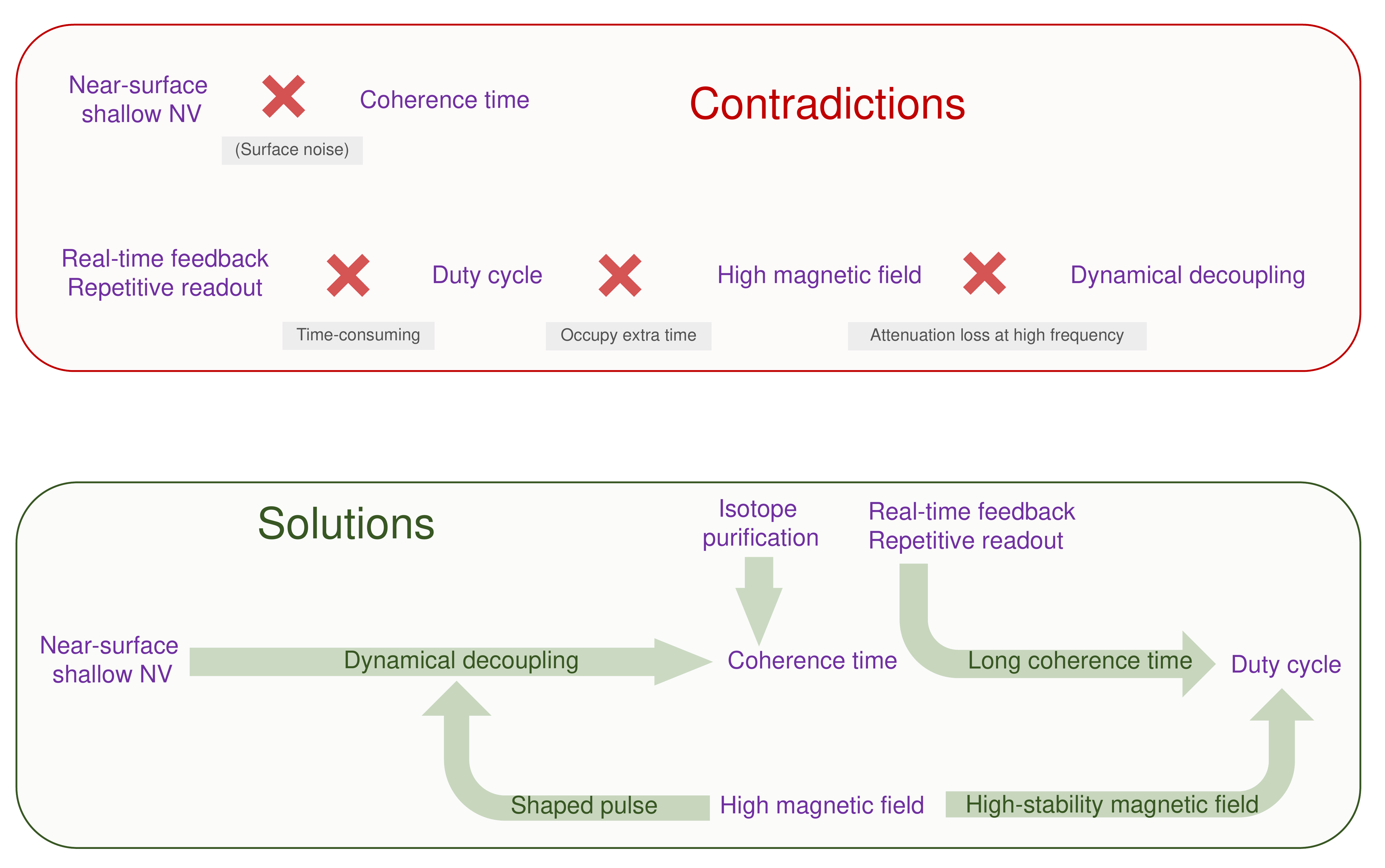}
\caption{\label{solution}\textbf{Contradictions between multiple techniques and solutions.}  When different technologies are combined together directly, a series of problems arise and the diagram shows our corresponding solutions.
}
\end{figure}

\begin{figure}
\centering
\includegraphics[width=1.0\textwidth]{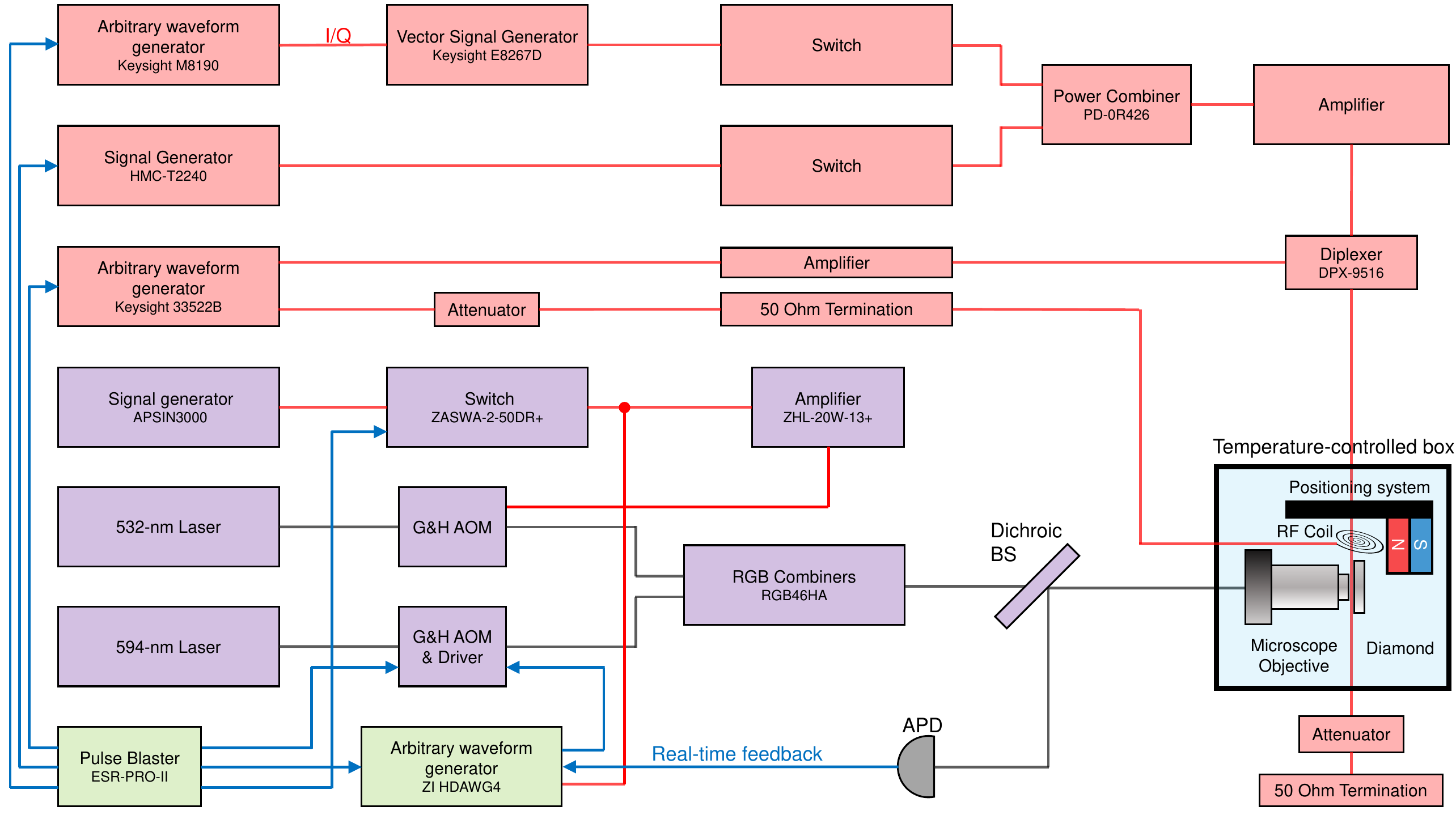}
\caption{\label{setup}\textbf{Simplified experimental setup.}  The setup mainly consists of microwave (MW) and radio-frequency (RF) circuits (red), optical systems (purple), synchronization and real-time feedback system (green), and a diamond platform inside a temperature-controlled box (blue). 
}
\end{figure}

\begin{figure}
\centering
\includegraphics[width=1.0\textwidth]{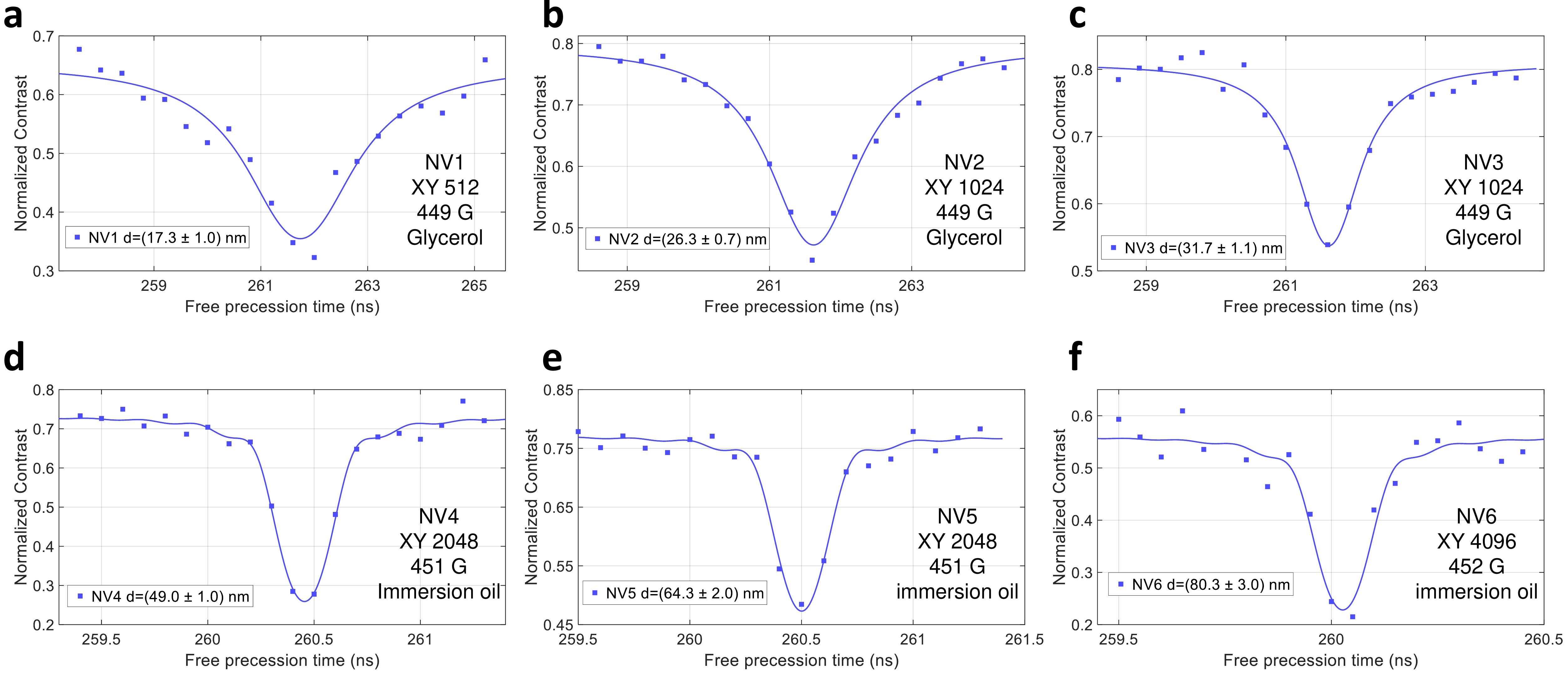}
\caption{\label{depth}\textbf{NMR proton spectra detected by NV centers.}  The NV number, the pulse sequence, the applied static magnetic field and the sample are given in the figures. The extracted NV depths are given in the symbol key.
}
\end{figure}

\begin{figure}
\centering
\includegraphics[width=1.\textwidth]{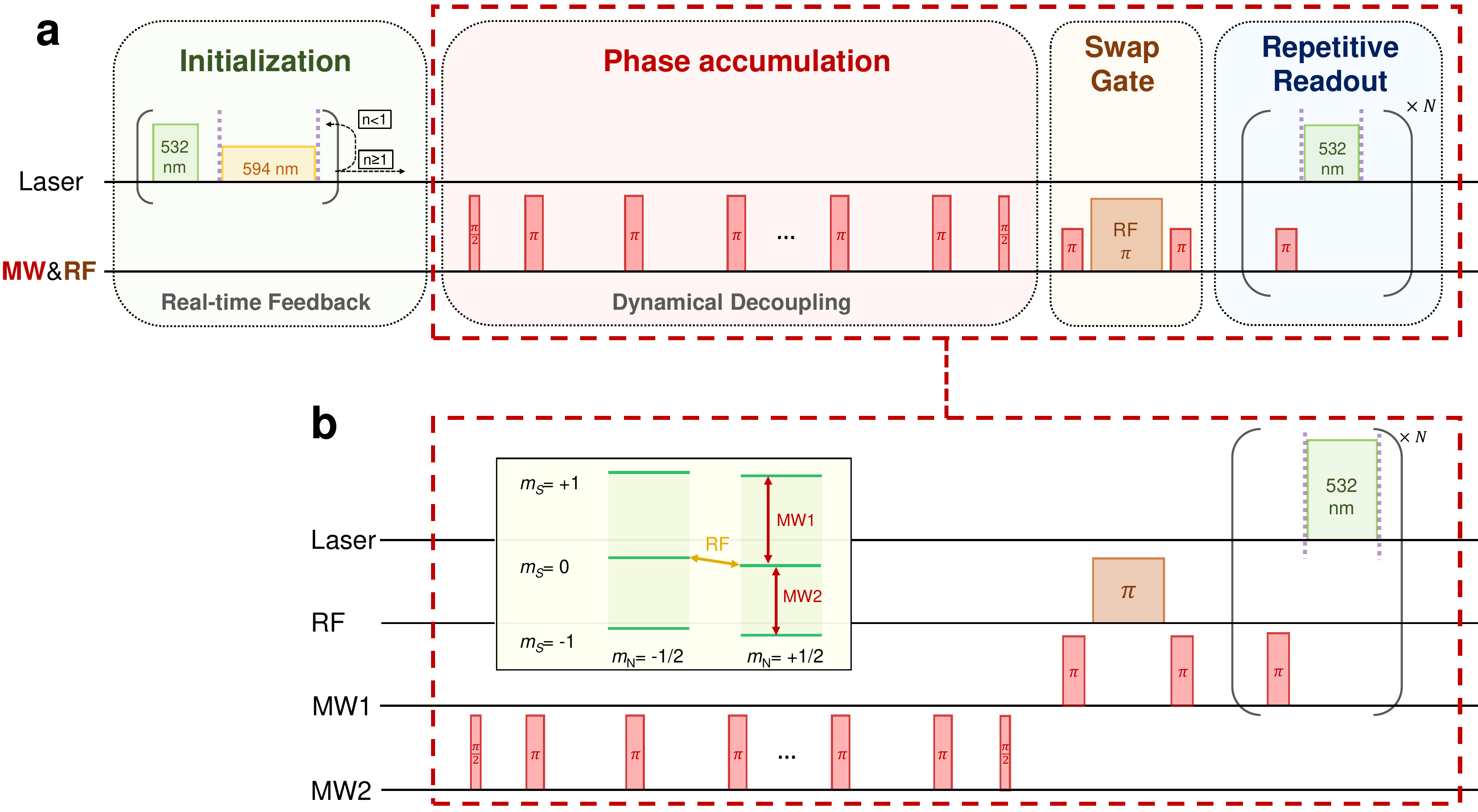}
\caption{\label{sequence}\textbf{Details of the interference sequecne.}  
(a) The interference sequecne for magnetic field measurement (Fig. 2a in the main text).
(b) The details of RF and MW control for the sequence in (a) and the corresponding energy levels of our NV-$^{15}$N system for illustration. The frequencies of RF, MW1 and MW2 are 3.305 MHz, 18.6 GHz and 24.3 GHz, respectively.
}
\end{figure}

\begin{figure}
\centering
\includegraphics[width=1.0\textwidth]{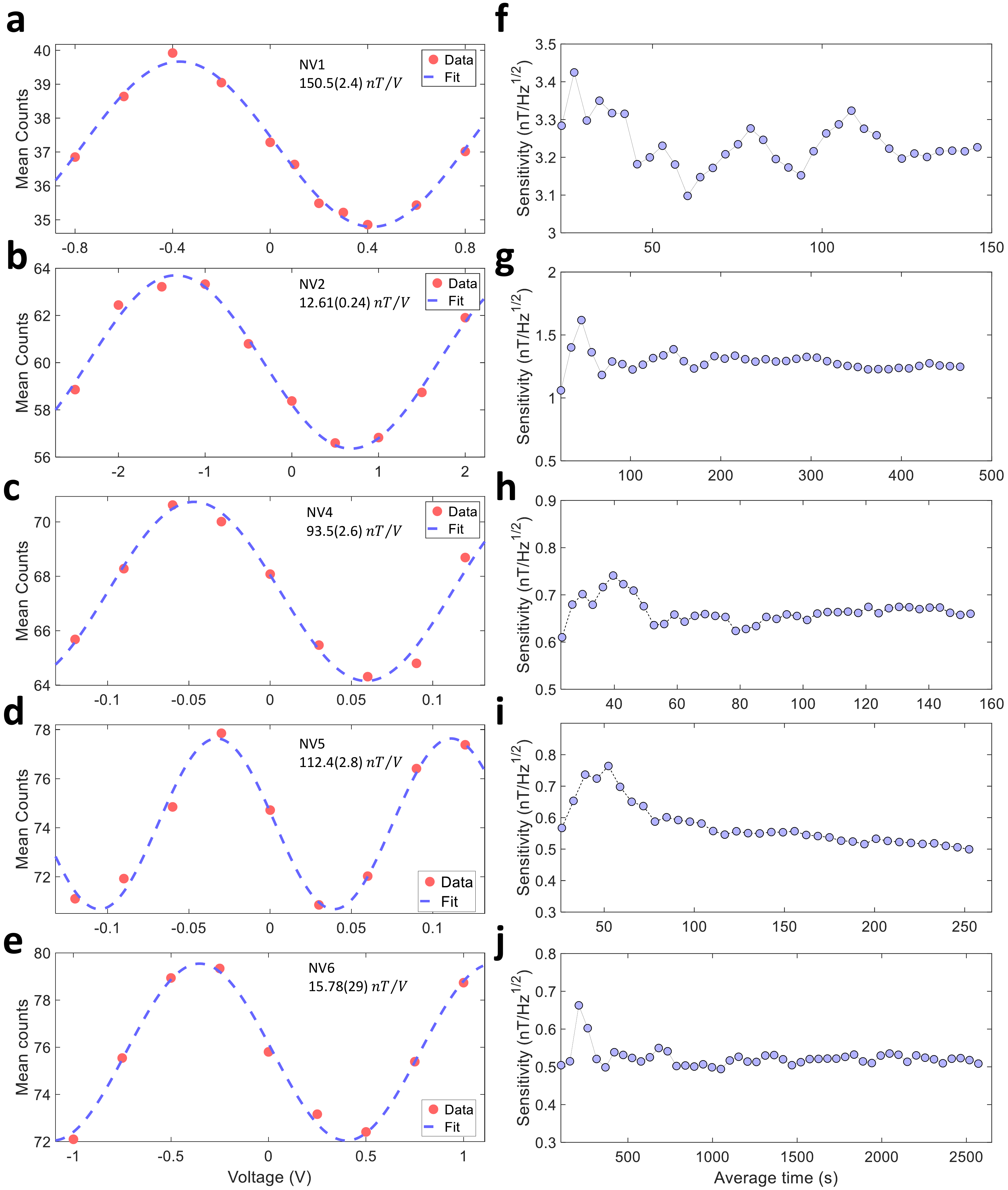}
\caption{\label{sensitivity}\textbf{Magnetic field measurement.}  
(a)-(e) The interference patterns for magnetic sensing. 
The dashed lines are given by fitting the experimental data.
(f)-(j) The magnetic sensitivities for five NV centers measured as a function of the average times.
}
\end{figure}

\begin{figure}
\centering
\includegraphics[width=1.0\textwidth]{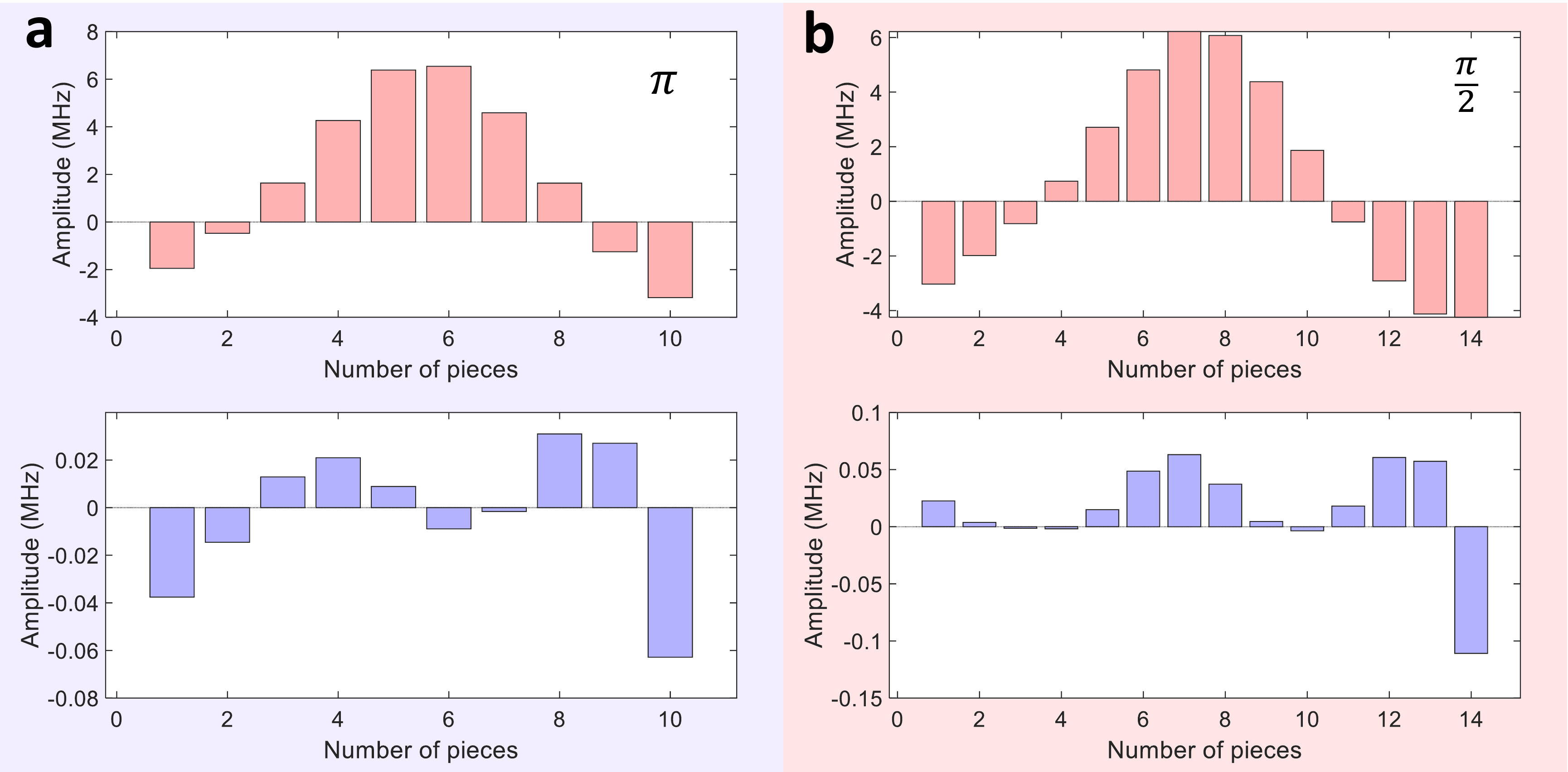}
\caption{\label{grape}\textbf{ Optimal control. }  
(a) The pulse sequence of the real (upper) and imaginary (lower) parts of the $\pi$ pulses used in dynamical decoupling sequences. The amplitude for each piece is denoted by Rabi frequency.  Each piece lasts for 25 ns with 10 pieces in total.
(b) The pulse sequence of the real (upper) and imaginary (lower) parts of the $\pi/2$ pulses. Each piece lasts for 25 ns with 14 pieces in total.
}
\end{figure}

\begin{figure}
\centering
\includegraphics[width=1.0\textwidth]{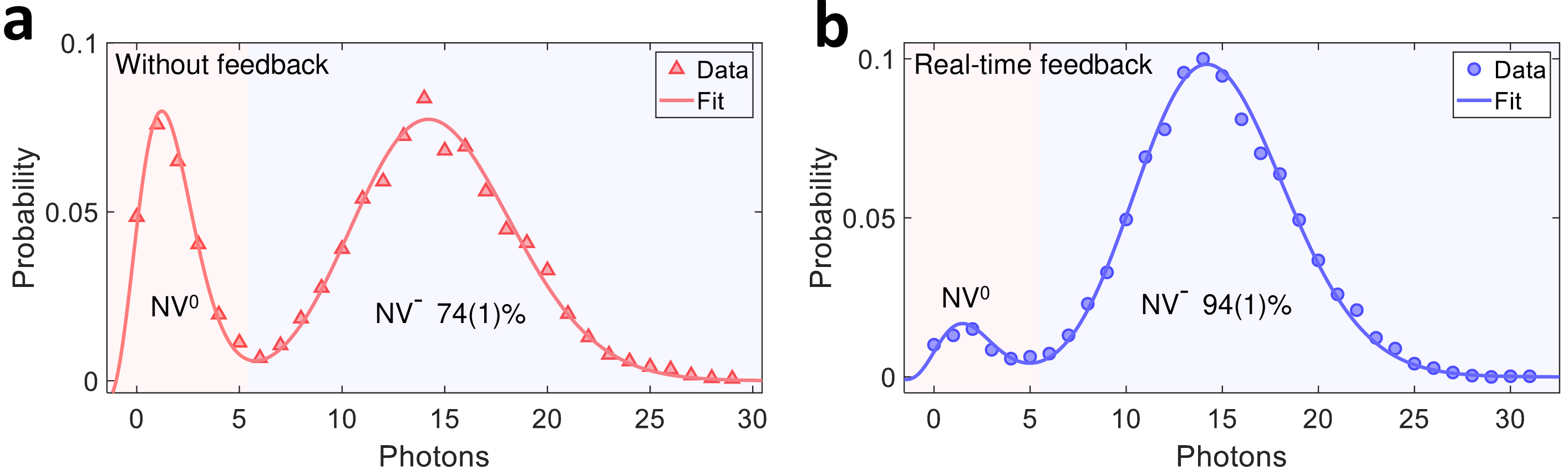}
\caption{\label{charge}\textbf{Real-time feedback for NV$^-$ preparation.}  
 (a) The distribution of two NV charge states (NV$^-$ and NV$^0$) after initialization without real-time feedback (red triangles, 74\% NV$^-$). The solid lines are two-peak Poissonian fitting curves for the data points.
 (b) The distribution of two NV charge states after real-time-feedback initialization (blue circles, 94\% NV$^-$).
}
\end{figure}

\begin{figure}
\centering
\includegraphics[width=0.8\textwidth]{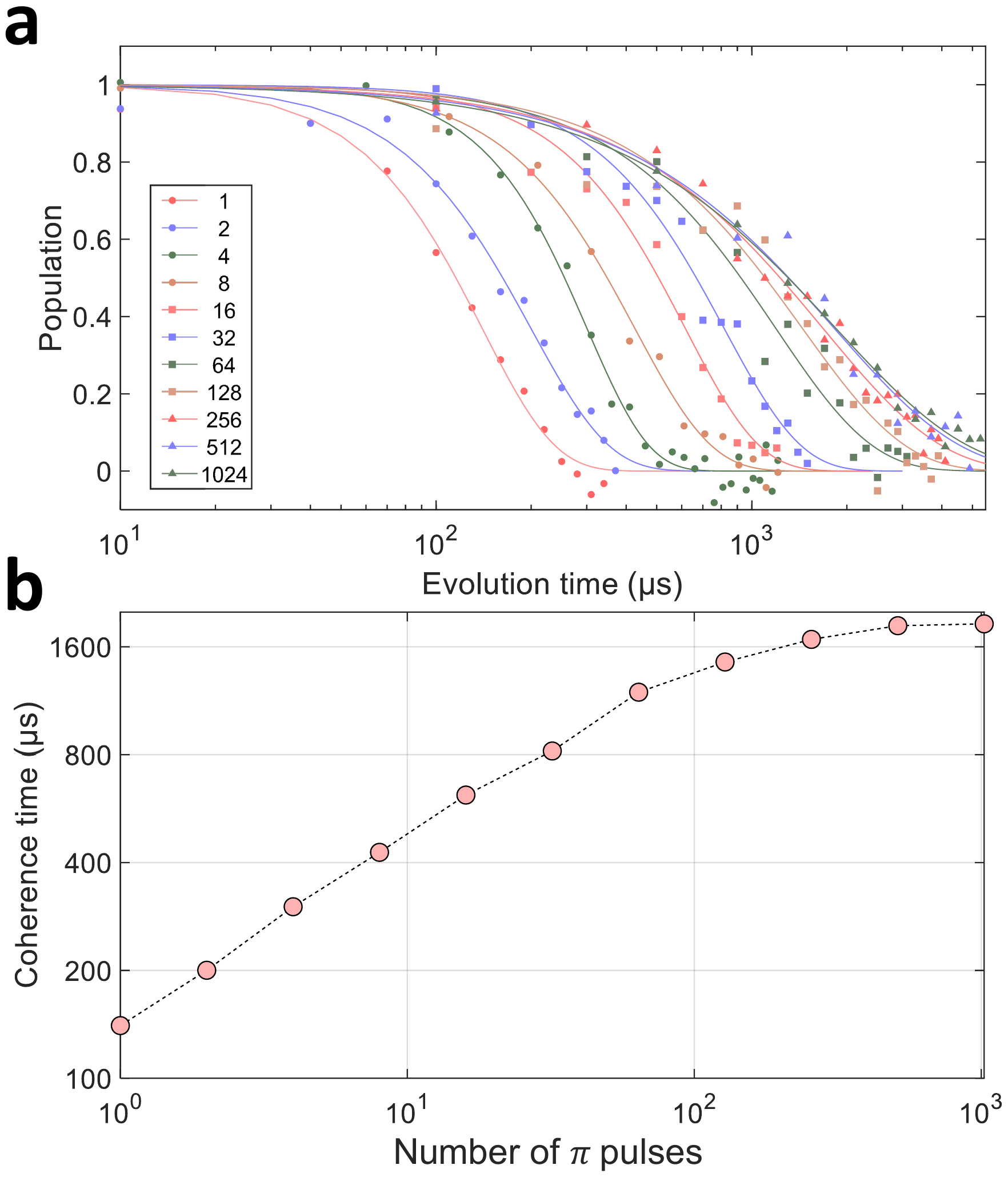}
\caption{\label{t2}\textbf{Decoherence behaviors.}  
(a) Decoherence behaviors for the NV center with the best energy resolution under multiple dynamical decoupling sequences with different numbers of $\pi$ pulses. The solid lines are the fittings with stretched exponential functions ($\exp(-(t/T_2)^p)$).
(b) Coherence times as a function of the number of $\pi$ pulses of dynamical decoupling sequences.
}
\end{figure}

\begin{figure}
\centering
\includegraphics[width=0.8\textwidth]{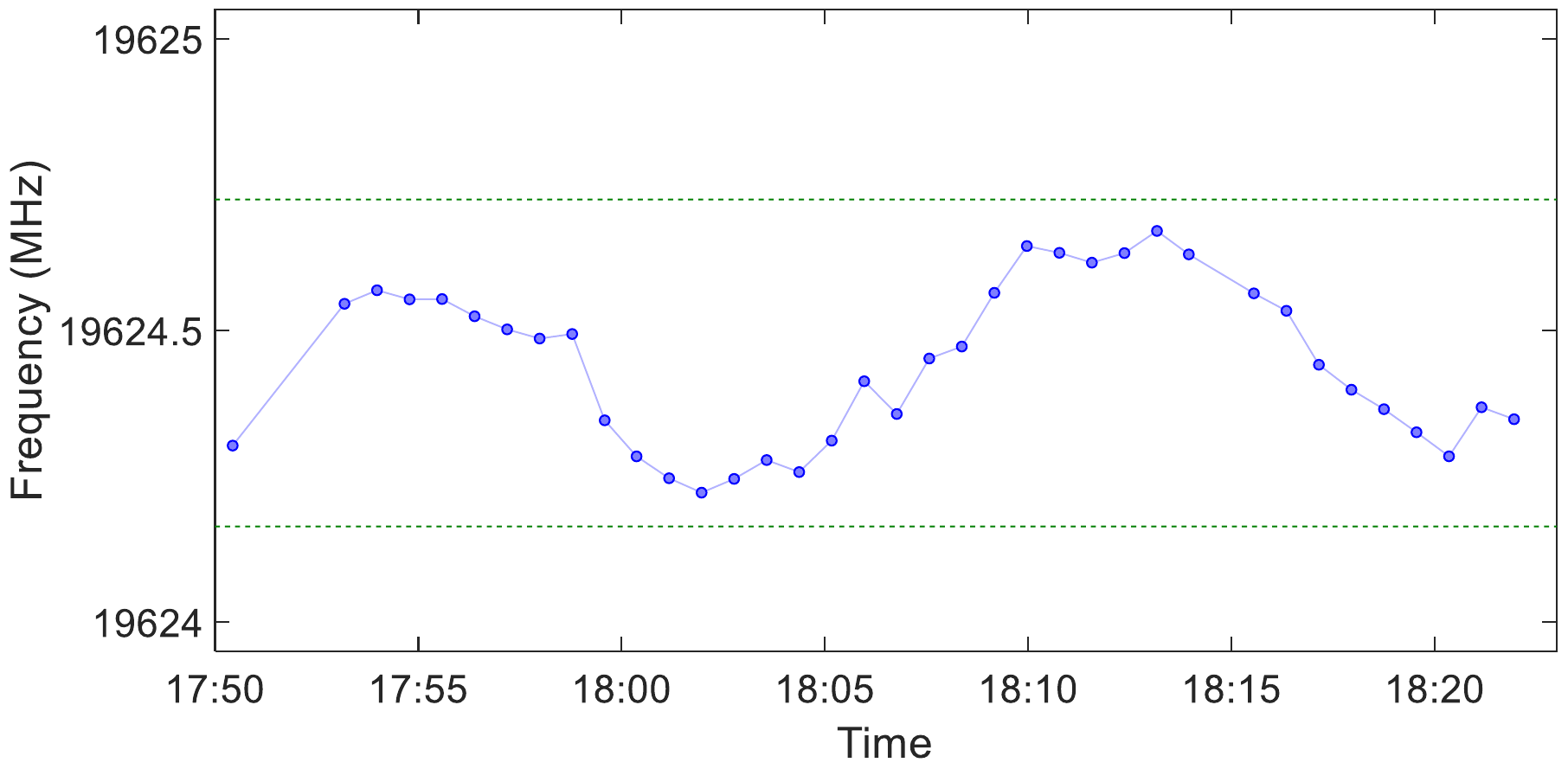}
\caption{\label{t2}\textbf{Magnetic field stability.}  
The blue line is the shift of the resonance spectral peaks during the experiments, which shows that the bias field oscillates within ~0.1 G (the green dashed lines).
}
\end{figure}

\begin{figure}
\centering
\includegraphics[width=1.0\textwidth]{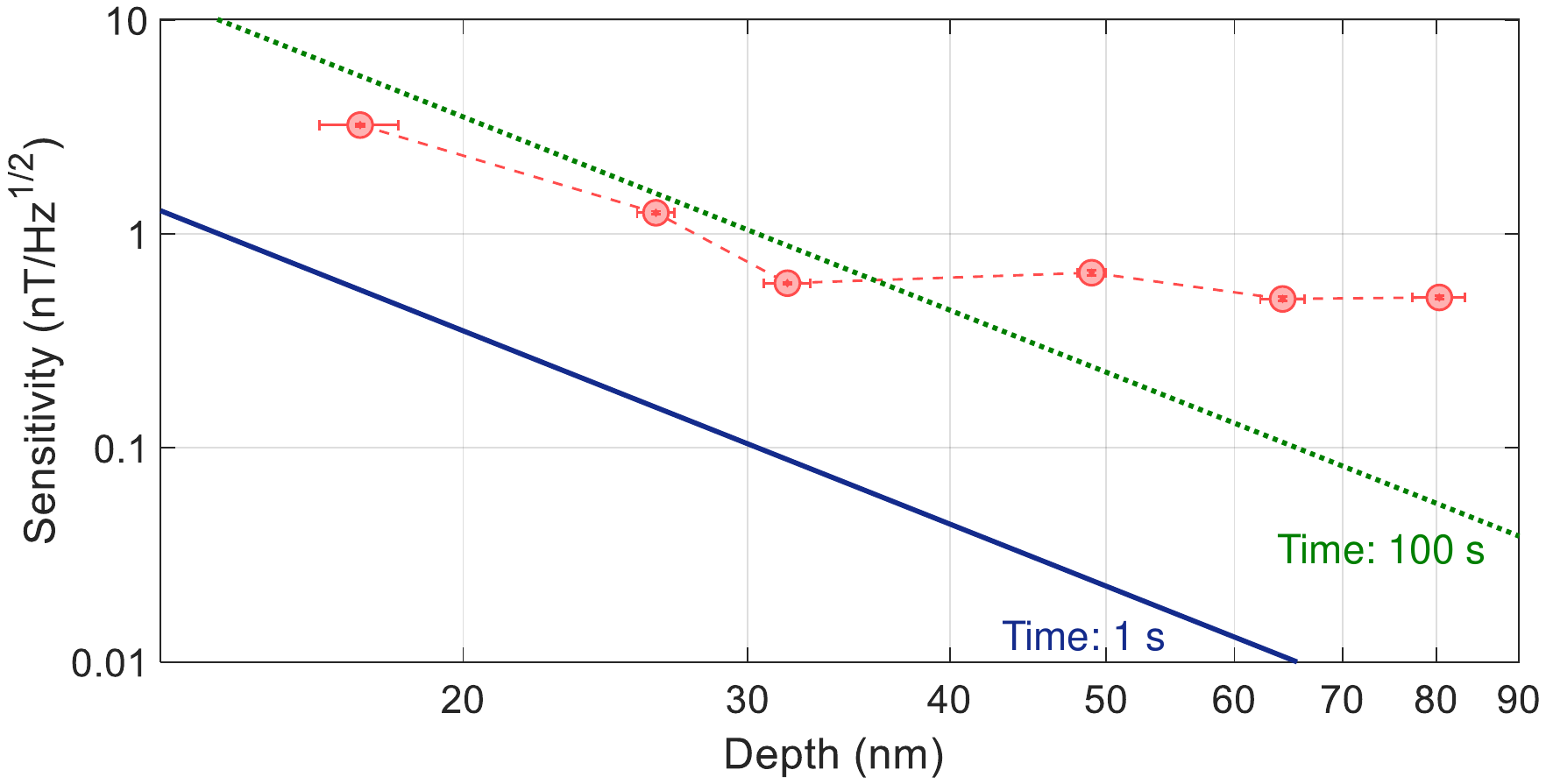}
\caption{\label{proton}\textbf{Sensitivity and the signal of a single proton.}  Measured sensitivities for six near-surface NV centers with different depths. The sensitivity needed for detecting a single proton (signal-to-noise ratio of 1) with 1 s (100 s) of data accumulation is calculated as a function of NV depth. For the NV center with the best energy resolution, it takes about 45 s to detect a single proton.
}
\end{figure}

\clearpage

\begin{figure}
\centering
\includegraphics[width=1.0\textwidth]{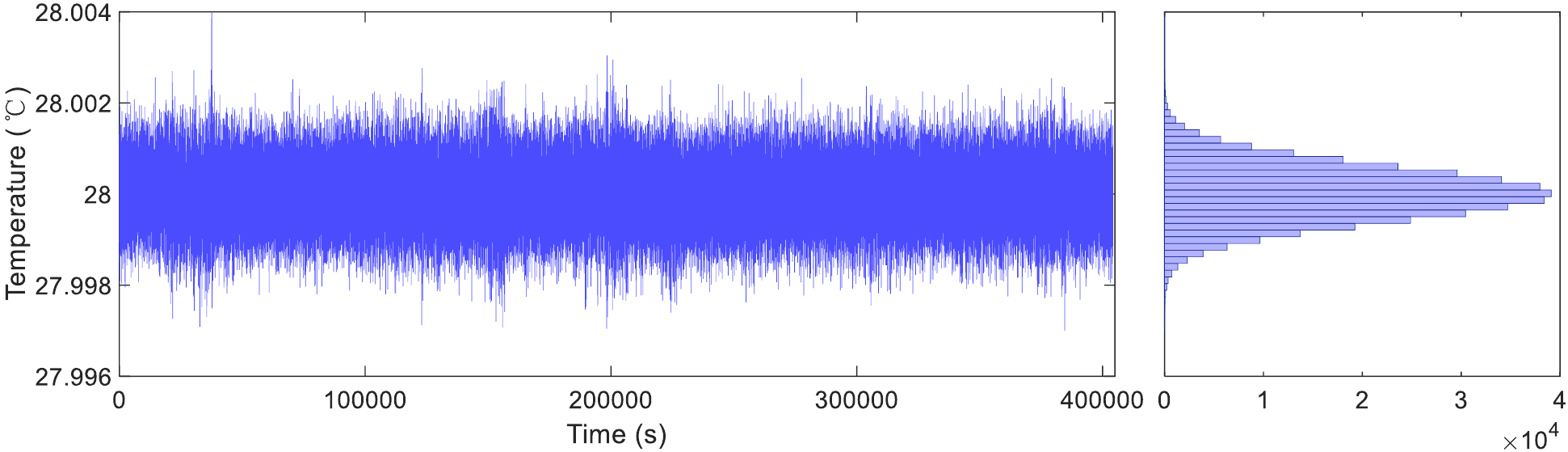}
\caption{\label{temperature}\textbf{Temperature stability.}  Time trace of the temperature inside the copper box. The fluctuation is 0.6 mK (one standard deviation).
}
\end{figure}